\documentclass[preprint,prd,preprintnumbers,amssymb,superscriptaddress,nofootinbib]{revtex4}
\pdfoutput=1
\usepackage{epsfig,psfrag,graphics,verbatim}
\usepackage[english]{babel}
\usepackage{graphicx}% Include figure files
\usepackage{dcolumn}% Align table columns on decimal point
\usepackage{bm}% bold math
\usepackage{slashed}
\usepackage{color}
\usepackage{amsmath}
\usepackage[]{units}
\usepackage{ulem}
\usepackage{multirow}
\usepackage{soul}

\newcommand{\DM}{\text{DM}}

\begin{document}
\preprint{ULB-TH/15-14}
\preprint{TUM-HEP 1006/15}

\title{Gamma-rays from Heavy Minimal Dark Matter
}

\author{Camilo  Garcia-Cely }
\affiliation{Service de Physique Th\'eorique, CP225, Universit\'e Libre de Bruxelles, Bld du Triomphe, 1050 Brussels, Belgium}

\author{Alejandro Ibarra}
\affiliation{Physik-Department T30d, Technische Universit\"at M\"unchen, James-Franck-Stra\ss{}e, D-85748 Garching, Germany}

\author{Anna S.~Lamperstorfer}
\affiliation{Physik-Department T30d, Technische Universit\"at M\"unchen, James-Franck-Stra\ss{}e, D-85748 Garching, Germany}
\author{Michel H.G. Tytgat}
\affiliation{Service de Physique Th\'eorique, CP225, Universit\'e Libre de Bruxelles, Bld du Triomphe, 1050 Brussels, Belgium}

\date{\today}

\begin{abstract}
Motivated by the Minimal Dark Matter scenario, we consider the annihilation into gamma rays of candidates in the fermionic 5-plet  and scalar 7-plet representations of $SU(2)_L$, taking into account both the Sommerfeld effect and the internal bremsstrahlung. Assuming the Einasto profile, we show that  present measurements of the Galactic Center by the H.E.S.S. instrument exclude the 5-plet and 7-plet as the dominant form of dark matter for masses between 1 TeV and 20 TeV, in particular, the 5-plet mass leading to the observed dark matter density via thermal freeze-out. We also discuss prospects for the upcoming Cherenkov Telescope Array, which will be able to probe even heavier dark matter masses, including the scenario where  the scalar 7-plet is thermally produced.
\end{abstract}

\maketitle

\section{Introduction}
\label{sec:intro}

While there is solid gravitational evidence for the existence of dark matter (DM), its actual nature still eludes us. The most popular hypothesis is that the DM is  made up  of massive and stable thermal relics of the Early Universe \cite{Lee:1977ua}, a scenario that points to DM particles with weak interactions. The existence of such DM candidates is predicted by many extensions of the Standard Model (SM) of particle physics, like supersymmetry \cite{Goldberg:1983nd,Ellis:1983ew,Jungman:1995df}. However, in recent years, a mixture of observational and experimental anomalies, together with theoretical considerations and the lack of supersymmetric signals at colliders, has lead to a complementary, more bottom-up approach. Among the plethora of models that have been discussed in the literature, some are characterized by their simplicity, as well as by  their rich phenomenology.
A non-exhaustive list includes singlet scalar \cite{McDonald:2001vt,Burgess:2000yq,Patt:2006fw} candidates, the inert doublet \cite{Deshpande:1977rw,Barbieri:2006dq,LopezHonorez:2006grw} and the so-called Minimal Dark Matter \cite{Cirelli:2005uq,Cirelli:2007xd,Hambye:2009pw}.

 Minimal Dark Matter, which we take as a motivation for the present work, is a particularly predictive framework that extends the SM with just one extra weak multiplet of particles, one of which is neutral and is a DM candidate. Assuming that the multiplet interacts dominantly through gauge interactions, 
 the only unknown parameter for a given multiplet is its mass, which may be fixed if one assumes that the DM was thermally produced in the Early Universe. In this paper we will take a more agnostic attitude and we will not make any assumption about the DM production mechanism but assume that the mass lies in the TeV range.

Of particular interest are the higher multiplets. In the original proposal it was pointed out that a fermionic 5-plet and a scalar 7-plet are the highest multiplets one may consider that are stable and are compatible with the Landau pole condition ~\cite{Cirelli:2005uq, Hamada:2015bra}. While it was pointed out in~\cite{DiLuzio:2015oha} that the 7-plet  may be unstable\footnote{The stability of the fermionic 5-plet is guaranteed by the fact that it has no dimension-5 effective couplings to the SM particles. The scalar 7-plet however has such a dimension-5 coupling, suppressed by a scale $\Lambda$, in which three scalar 7-plets in a symmetric triplet state couple to two SM scalar doublets. This term leads at one loop to the decay of the  DM candidate into SM particles, the lifetime being ${\cal O}(10^{-8}$ s$)$ for $\Lambda \sim 10^{15}$ GeV and $M_\text{DM} \sim 1$ TeV \cite{DiLuzio:2015oha}. To prevent such coupling, an  extra discrete symmetry must be postulated.}, we use these two models as benchmark scenarios to analyze gamma-ray signals of DM particles belonging to large SU(2) multiplets, with spin 0 and with spin 1/2. Thermal relics from  higher multiplets tend to have a very large mass, typically a few TeV, lying  beyond the reach of the LHC. Constraints from indirect searches are thus of greatest significance for such DM candidates.

In the present work we focus on the constraints on DM candidates in large $SU(2)_L$ representations, concretely the fermionic 5-plet and the scalar 7-plet, that stem from the non-observation of the gamma-rays which are hypothetically produced in their self-annihilation in the Milky Way Center.   DM candidates in smaller $SU(2)_L$ representations have been largely discussed in the recent literature, in particular the fermionic triplet, or wino DM, so our analysis is complementary to those of Refs.\cite{Chun:2012yt,Cohen:2013ama,Baumgart:2014vma,Ovanesyan:2014fwa,Baumgart:2014saa,Beneke:2014hja,Chun:2015mka}. The  motivation for studying higher multiplets, like the 5- and 7-plets, is manifold. First, we want to confront these DM candidates to the most recent data from the High Energy Spectroscopic System (H.E.S.S.) \cite{Abramowski:2013ax} and to the sensitivity of the future Cerenkov Telescope Array (CTA) \cite{Consortium:2010bc}, along the vein of the forecast analysis made in \cite{Ibarra:2015tya}. Second, as emphasized in \cite{,Cirelli:2005uq,Cirelli:2007xd}, since the DM mass is much greater than the electroweak scale, indirect signals from these candidates are subject to non-perturbative effects~\cite{Hisano:2004ds,Hisano:2002fk, Cirelli:2007xd}. The computation of the non-perturbative effects is plagued with numerical instabilities~\cite{Cohen:2013ama}, which become even more acute for higher multiplets. For our work, and in order to obtain reliable results, we have developed a new algorithm which cures the numerical instabilities. Lastly, we also include in our analysis the annihilations into $W^+W^-\gamma$, a process dubbed internal bremsstrahlung, which generates a sharp spectral feature in the gamma-ray energy spectrum \cite{Bergstrom:1989jr,Beacom:2004pe,Boehm:2006df} (see also \cite{Garcia-Cely:2013zga}). 
Including the two-to-three annihilation increases the flux into sharp spectral features and, as we will argue, even provides, for certain DM masses, the dominant contribution to the total gamma-ray spectrum, due to cancellations among the amplitudes contributing to other processes.

The rest of our article is organized as follows. In Section~\ref{sec:MDM} we briefly review the properties of the fermionic 5-plet and scalar 7-plet DM scenarios which are relevant for our analysis. In Section~\ref{sec:SE} we describe our procedure to evaluate the non-perturbative effects on the annihilation process, and in Section~\ref{sec:cross} we derive the cross sections for the relevant annihilation channels. Then, in Section~\ref{sec:HESS} we calculate constraints on the model using  H.E.S.S data from the Galactic Center and, in Section~\ref{sec:CTA}, a forecast for the CTA reach.  Lastly, we present our conclusions in Section~\ref{sec:conclusions}. We also include an Appendix describing our algorithm to numerically calculate the Sommerfeld enhancement factors. 

\section{The 5-plet   and 7-plet DM  candidates}
\label{sec:MDM}

We extend the SM particle content by one extra fermionic or scalar $SU(2)_L$ multiplet, motivated by the Minimal Dark Matter framework~\cite{Cirelli:2005uq} (see also~\cite{Cirelli:2009uv} for a summary). Then, the Lagrangian reads
\begin{equation}
{\cal L} = {\cal L}_\text{SM} + {1\over 2} \bar \chi (i \slashed{D} - M) \chi \qquad \mbox{(fermion)}
\end{equation}
or 
\begin{equation}
{\cal L} = {\cal L}_\text{SM} + {1\over 2} \left(\vert D_\mu \chi\vert^2-  M^2 \vert\chi\vert^2\right) \qquad \mbox{(scalar)}\;.
\end{equation}
In this work we focus on the fermionic 5-plet and scalar 7-plet as benchmarks for large multiplets. Both have zero hypercharge $Y=0$, so that the 5-plet is Majorana and the 7-plet is real.  Then,  each of those multiplets can be written in terms of their components as
\begin{equation}
\chi= \begin{pmatrix} \DM^{2+}\\ \DM^+\\ \DM \\ -\DM^-\\ \DM^{2-}\end{pmatrix} \text{    for the 5-plet},\,
\hspace{20pt}
\chi= \begin{pmatrix} \DM^{3+}\\ \DM^{2+}\\ \DM^+\\ \DM \\ -\DM^-\\ \DM^{2-} \\ -\DM^{3-}\end{pmatrix}\text{        for the 7-plet.}\,
\end{equation}
Here the relative signs have been introduced in order to make the multiplets isospin self-conjugate. The notation for the multiplet components is self-explanatory. 

Electroweak symmetry breaking induces a mass splitting between the different particles in the multiplet. Both for scalars and fermions, weak isospin breaking effects necessarily arise  at the one-loop level, through the exchange of $Z$ or $W$ gauge bosons \cite{Cirelli:2005uq},
generating a mass splitting between the particles of charge $Q$ and $Q'$ which is given by
\begin{equation}
M_Q - M_{Q'} \approx (Q^2 - Q^{\prime 2})\, \Delta\,,\hspace{30pt}\text{where}\hspace{30pt}\Delta \equiv \alpha_2 \sin^2\left({\theta_W\over 2}\right) M_W \approx 166 \text{ MeV}.
\end{equation}
(Note that $\Delta$ corresponds to  the  mass difference  between the neutral and the charge $\pm 1$ states.) As a result,  
the neutral particle is the lightest component of the multiplet. For the fermionic 5-plet, the dominant source of isospin breaking is the above-mentioned quantum effect, since the isospin breaking effects induced by possibly existing non-renormalizable operators are naturally suppressed. 
This is not  a priori the case for scalar candidates, which may have renormalizable couplings to the SM scalar doublet~\cite{Cirelli:2005uq,Hambye:2009pw}. The relevant coupling is
\begin{equation}
{\cal L} \supset  \lambda\;  \left(\chi^\dagger T^a \chi\right) \; \left(H^\dagger \tau^a H\right)\;,
\end{equation} 
where $H$ is the SM scalar doublet and $T^a$ are the generators of the $SU(2)_L$ algebra. Nevertheless, $\left(\chi^\dagger T^a \chi\right)$ is identically vanishing for real scalars \cite{Hambye:2009pw}, which is the case we are considering here, therefore the splitting for the scalar 7-plet is naturally dominated by loop corrections, as in the fermionic case. 

If the fermionic 5-plet or scalar 7-plet candidate is the dominant form of DM, then the only unknown parameter for a given multiplet is the  DM mass, which may be fixed by requiring that the DM is generated via thermal freeze-out. Taking into account the Sommerfeld enhancement in the Early Universe, this requirement leads to a 5-plet mass of about 10 TeV and a 7-plet mass of about 25 TeV \cite{Cirelli:2007xd}. Nevertheless, other scenarios are also possible. For instance, for  masses below the above mentioned values, thermal freeze-out of  DM particles only produces a fraction of the total DM density, therefore other DM species would be necessary to explain the astrophysical and cosmological data (see, {\it e.g.} \cite{Heeck:2015qra}). Yet, annihilations of DM particles in the Galactic Center could generate observable signals. On the other hand, for  masses above those values, thermal freeze-out overproduces DM particles. However, some dilution mechanism could be at work reducing the DM density (see for instance \cite{Kane:2015qea}  for a discussion of some possible alternative non-thermal histories). Besides,  DM particles might have never reached thermal equilibrium, as it is the case in cosmological scenarios with a reheating temperature below the DM mass. In this case, some non-thermal mechanism should be postulated to generate the observed DM abundance (or a fraction of it). Given our ignorance of the dark sector and of the thermal history of the Universe, we will leave both the DM mass and the DM density today as free parameters, and we will concentrate on the indirect detection of this class of scenarios. In particular we will concentrate on the constraints on DM annihilation into gamma-rays from observations of the Milky Way Center, most notably from searches for sharp spectral features. Since the annihilation cross sections are fixed for a given DM mass, the non-observation of a gamma-ray excess will lead to constraints on the density fraction of the given  DM candidate, regardless of any hypothesis about the DM production mechanism. 

In the next section we will discuss in more detail our procedure to calculate the non-perturbative effects in the annihilation in the Milky Way Center of the fermionic 5-plet or the scalar 7-plet and which play a central role in correctly assessing the prospects for indirect detection of  these DM candidates.

%%%%%
\section{The Sommerfeld Effect}
\label{sec:SE}

\begin{table}[b!]
\centering
\begin{tabular}{c|c|c|}\cline{2-3}
 &\multirow{2}{*}{ 5-plet} & \multirow{2}{*}{ 7-plet} \\
 & & 
\\\cline{1-3}
\multicolumn{1}{|c|}{
\multirow{2}{*}{$V(r)$}
} & $ \begin{pmatrix} 4\,C & -2\,B  & 0 \\ -2\,B  & \,C & -3\,\sqrt2\,B  \\ 0 & -3\,\sqrt2\,B   & 0\end{pmatrix}$
&$\begin{pmatrix}  9\,C & -3\,B  & 0 & 0 \\ -3\,B  & 4\,C & -5\,B  & 0 \\ 0 & -5\,B  &  \,C & -6\sqrt2 \,B  \\ 0 &  0 & -6\sqrt2  \,B   & 0\end{pmatrix}$\\\cline{2-3}
\multicolumn{1}{|c|}{
\multirow{2}{*}{}
 }& \multicolumn{2}{c|}{
\multirow{2}{*}{$B=  \frac{g^2}{4\pi r} e^{-m_W r}$ \hspace{20pt} $C = 2\Delta - \frac{ e^2}{4\pi r}- \frac{g^2 c_W^2 }{4\pi r}\,e^{-m_Z r}$}
}
\\
\multicolumn{1}{|c|}{}  &  
\multicolumn{2}{c|}{
}\\\hline\hline
\multicolumn{1}{|c|}{
\multirow{2}{*}{${\cal W}_{W^+W^-} $}
}&
\multirow{2}{*}{ $\,\, g^2 \left(2\,\,5\,\,3\sqrt{2}\right) $ }  & 
\multirow{2}{*}{$ \,\, g^2 \left(3\,\, 8\,\,11\,\, 6\sqrt{2} \right) $} \\
\multicolumn{1}{|c|}{}& & 
 \\\cline{1-3}
\multicolumn{1}{|c|}{
\multirow{2}{*}{ ${\cal W}_{\gamma\gamma}$} 
}& 
\multirow{2}{*}{$\sqrt{2} e^2 (2^2\,\, 1 \,\,0)$ }& 
\multirow{2}{*}{$  \sqrt{2} e^2 (3^2\,\,2^2\,\,1\,\, 0)$} \\
\multicolumn{1}{|c|}{}& &
 \\\cline{1-3}
\multicolumn{1}{|c|}{
\multirow{2}{*}{${\cal W}_{Z Z} $}
} &
\multicolumn{2}{c|}{
\multirow{2}{*}{ $\frac{1}{\tan^2\theta_W} {\cal W}_{\gamma\gamma}$}
}   \\
\multicolumn{1}{|c|}{}& \multicolumn{2}{c|}{}  
 \\\cline{1-3}
\multicolumn{1}{|c|}{
 \multirow{2}{*}{${\cal W}_{\gamma Z}$ }
} &
\multicolumn{2}{c|}{
\multirow{2}{*}{$\frac{\sqrt2}{\tan\theta_W}\,{\cal W}_{\gamma\gamma}$}  
}   \\
\multicolumn{1}{|c|}{}& \multicolumn{2}{c|}{}  
  \\\hline
\end{tabular}
\caption{\small Potential matrices for the 5-plet and 7-plet representations, as well as row vectors ${\cal W}$ entering the calculation of the annihilation cross sections ({\it cf.} Eq.~\eqref{SEsigmav}).  The DM pair state vector is (DM$^{2+}$DM$^{2-}$,  DM$^{+}$DM$^{-}$, DM\,DM) for the 5-plet and (DM$^{3+}$DM$^{3-}$, DM$^{2+}$DM$^{2-}$,  DM$^{+}$DM$^{-}$, DM\,DM) for the 7-plet.}
\label{table:GandV}
\end{table}

 The fermionic 5-plet and the scalar 7-plet DM particles annihilate via the exchange of gauge bosons.  Since these are very light compared to the annihilating particles, the long range interaction can significantly distort the wave function of the non-relativistic particles involved in the annihilation. As a result, the perturbative calculation of the annihilation cross section breaks down and instead a non-perturbative approach ought to be pursued, leading  generically to a larger value of the cross section. This is the so-called Sommerfeld enhancement, first described in Ref.~\cite{Sommerfeld} for electron-nucleon scattering at low relative velocities. In this paper, we calculate this effect following closely  the formalism presented in Ref.~\cite{Hisano:2004ds} and \cite{Cirelli:2007xd}, consisting in first calculating the non-relativistic potential associated to the exchange of gauge bosons and then solving the corresponding Schr\"odinger equation for the two-state wave-functions involved in the annihilation. From the solution one obtains a set of non-perturbative enhancement factors, which are finally used to calculate the annihilation cross section in combination with the perturbative amplitudes.

For the 5-plet case, due to the exchange of gauge bosons, the dark matter pair states $\DM\, \DM$ mix with the states $\DM^+\,\DM^-$ and $\DM^{2+}\, \DM^{2-}$. Consequently, the matrix $g(r)$ introduced in Ref.~\cite{Hisano:2004ds}, and which encodes the Sommerfeld effect, is $3\times3$. For the 7-plet case, DM pairs additionally mix with the state $\DM^{3+}\, \DM^{3-}$, and the resulting $g(r)$ matrix is $4\times4$.  These matrices satisfy the Schr\"odinger equation~\cite{Hisano:2004ds}
\begin{eqnarray}
\frac{1}{M}g''(r) + \left(\dfrac{1}{4} M v^2 {1\!\!1} - V(r) \right) g(r) =0\,,
\label{SoDE}
\end{eqnarray}
where $v$ is the relative velocity of the $\DM$ particles, concretely $v=2\times 10^{-3}$ in our analysis, and $V(r)$ is a central potential, which includes the exchange of gauge bosons among pairs of particles and the mass splitting among them. We will consider the s-wave piece of the annihilation cross sections both for scalar and Majorana particles, therefore the potential matrices $V(r)$ for each $SU(2)_L$ representation are the same for scalars and for fermions; their explicit form is shown in Table \ref{table:GandV}.

Since the  DM relative kinetic energy in the Milky Way Center is smaller than the mass splitting between the DM and the charged particles of the multiplet, only DM pairs  can exist  at large distances in the scattering process.  The boundary conditions for the  Schr\"odinger equation Eq.~(\ref{SoDE}) then read 
\begin{eqnarray}
g(0)&=&{1\!\!1}\;, \nonumber \\
g(r) &\to& e^{(i Mv/2)\sqrt{{1\!\!1}-4V(\infty)/(Mv^2)}}
D = \ e^{ ir Mv/2 }\begin{pmatrix}
  0   &  \cdots & 0 \\
  0   &  \cdots & 0 \\
\vdots & \ddots & \vdots \\
 \cdots & d_{+-} & d_{00}
 \end{pmatrix}
\text{  when   }r\to \infty\,.
\label{BC}
\end{eqnarray}
Here $D$ is a constant matrix and $d \equiv (\cdots \,\,d_{+-}\,\, d_{00})$ is the only non-zero row of the matrix $g(r)$ after factorizing out the out-going wave describing the DM pairs at infinity. Notice that when gauge interactions are turned-off, $d$ is just a unit vector. Consequently, its components can be interpreted as non-perturbative  enhancement factors due to the exchange of gauge bosons. These components, furthermore, are all real up to a global phase.\footnote{This result follows from the fact that the potential matrices $V(r)$ are not only Hermitian but also real. Then, using Eq.\eqref{SoDE}, it is possible to prove that the following matrices are conserved
\begin{eqnarray}
J_1 & =& g(r)^T\, g'(r) - g'(r)^T\, g(r)\,,\hspace{40pt}
J_2  = i\left(g(r)^\dagger\, g'(r) - g'(r)^\dagger\, g(r)\right)\,.\nonumber
\end{eqnarray}
Noticing that the boundary conditions Eq.~\eqref{BC} at infinity imply $J_1=0$ and $J_2 = M v \,d^\dagger d$, while at $r=0$  $ J_1 = g'(0)- g'(0)^T $ and $ J_2 = i(g'(0)- g'(0)^\dagger) $, one obtains that the matrix $J_2 = M v \,d^\dagger d$ must be real. Finally, since the row vector $d$ can in principle be complex, one concludes that the Sommerfeld factors must share a common global phase,  which is irrelevant for the evaluation of the annihilation cross sections, as follows from Eqs. \eqref{SEsigmav} or \eqref{CrossIBDiff}.  }

\begin{figure}[h!]
\begin{center}
\includegraphics[width=10.cm]{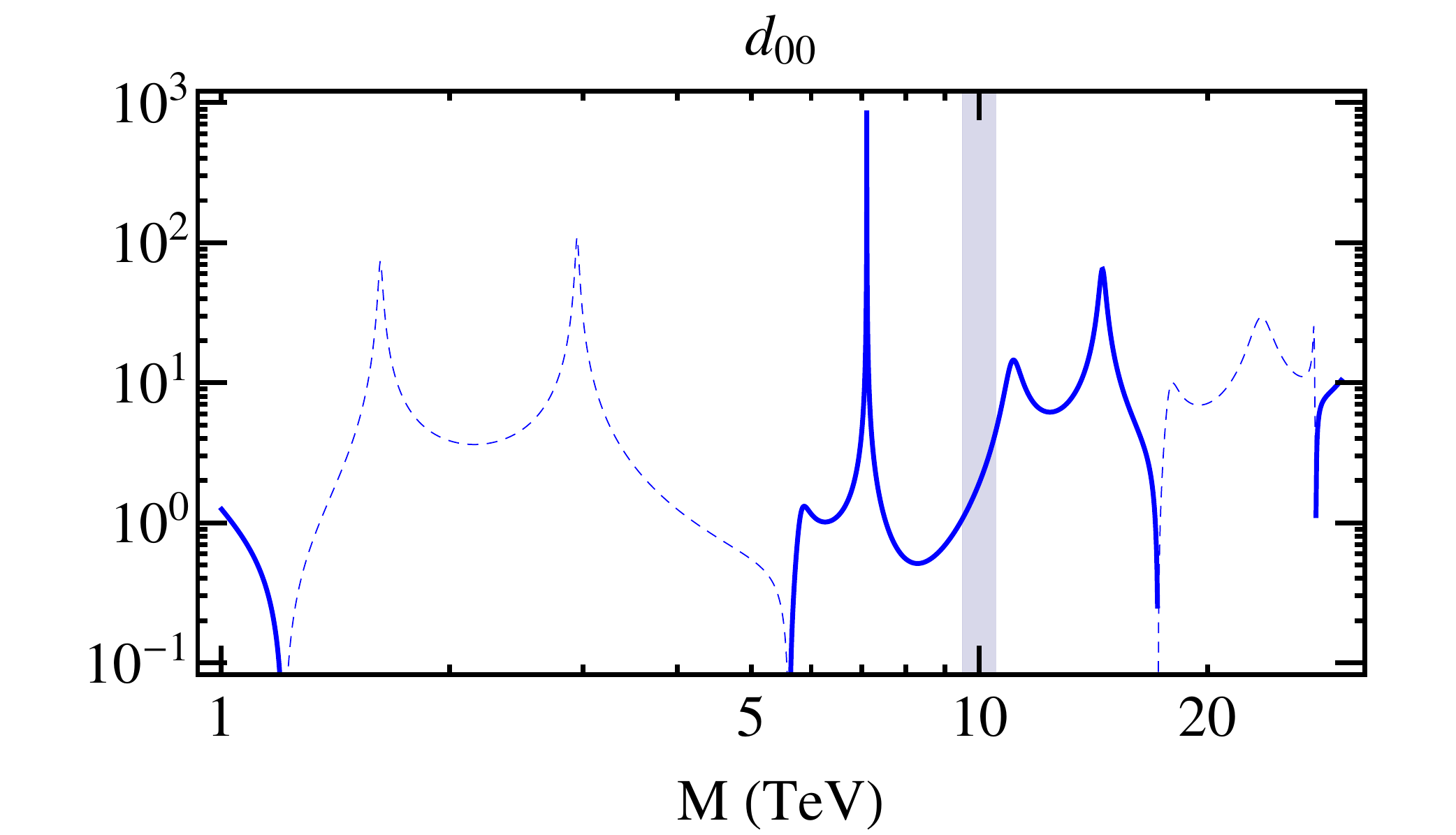}\\
\includegraphics[width=10.cm]{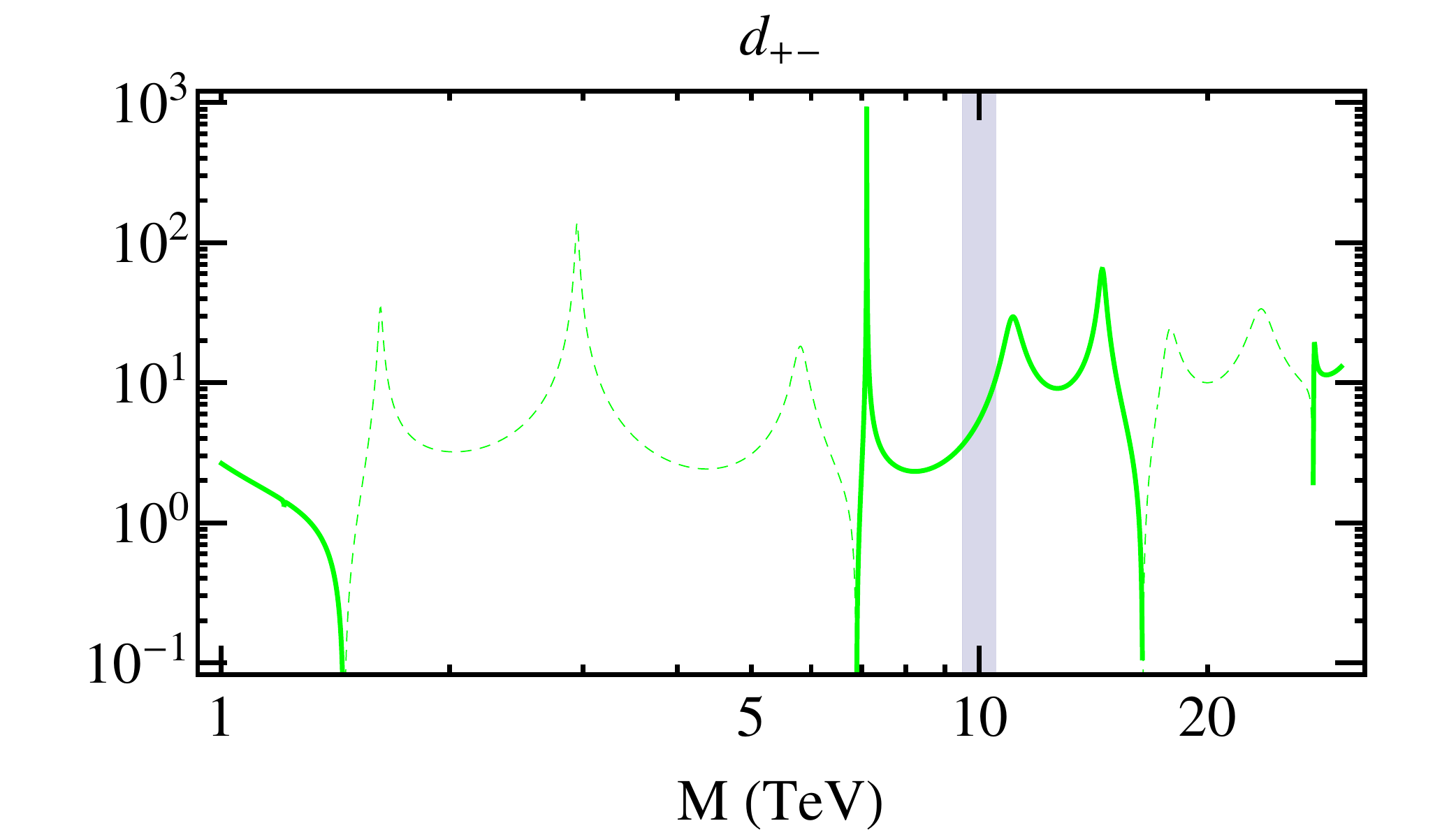}\\
\includegraphics[width=10.cm]{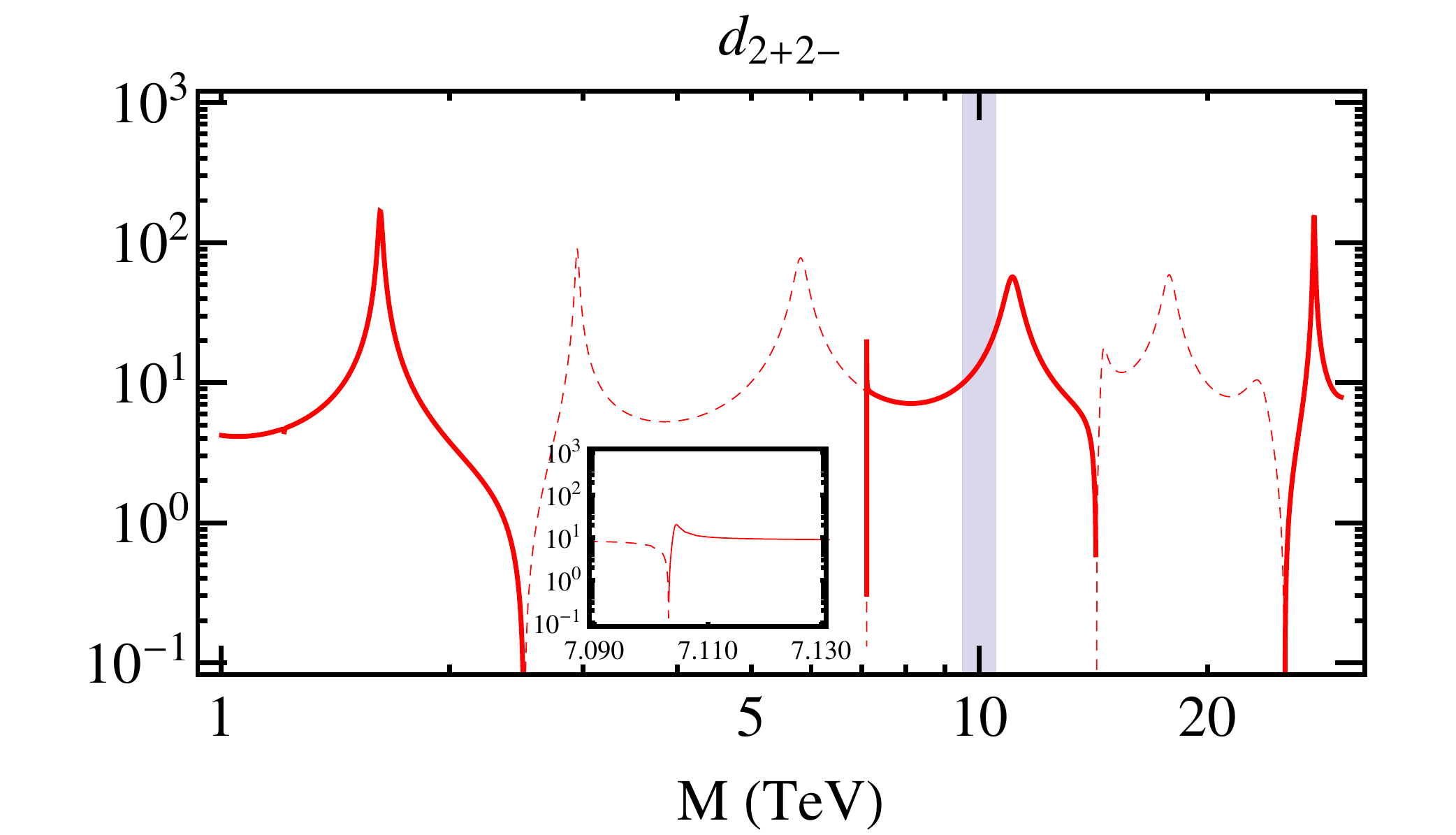}
\caption{Sommerfeld enhancement factors for the fermionic 5-plet. The solid (dashed) line corresponds to positive (negative) values of the enhancement factors. The gray band indicates the value of the mass reported in Ref.~\cite{Cirelli:2007xd} which leads to the observed DM abundance via thermal freeze-out.
}
\label{fig:Qd}
\end{center}
\end{figure}

\begin{figure}[h!]
\begin{center}
\includegraphics[width=12.5cm]{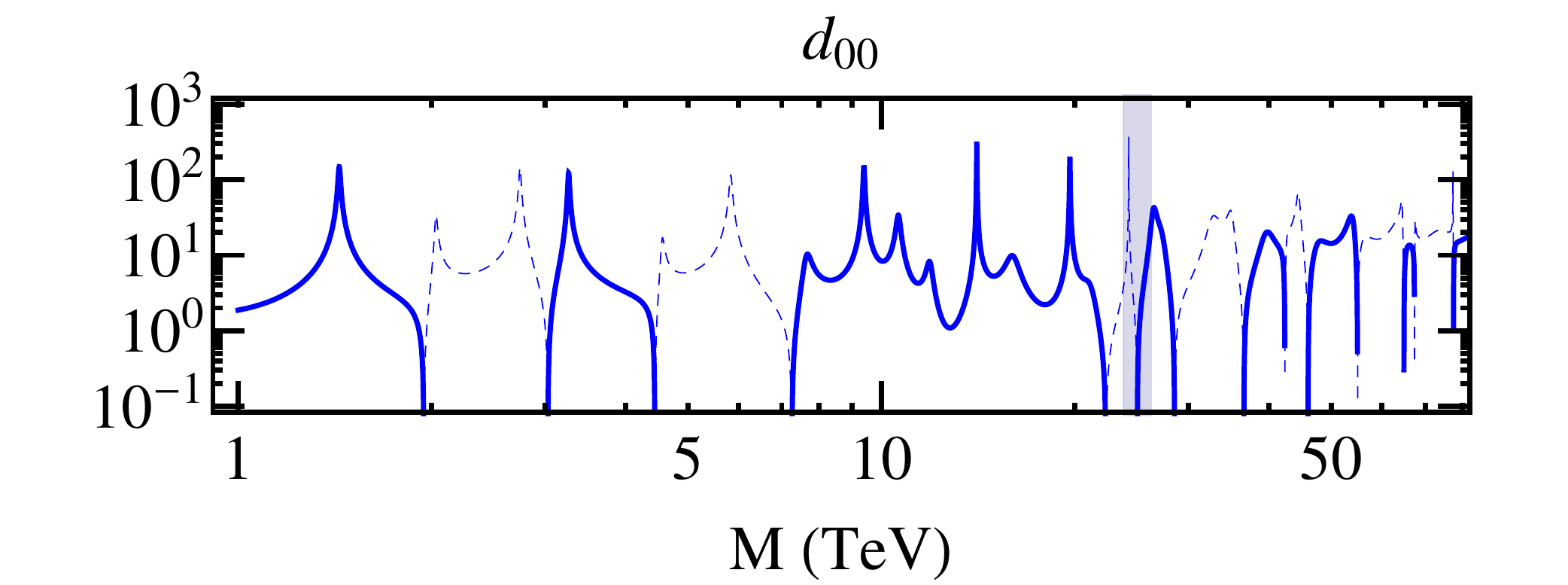}\\
\includegraphics[width=12.5cm]{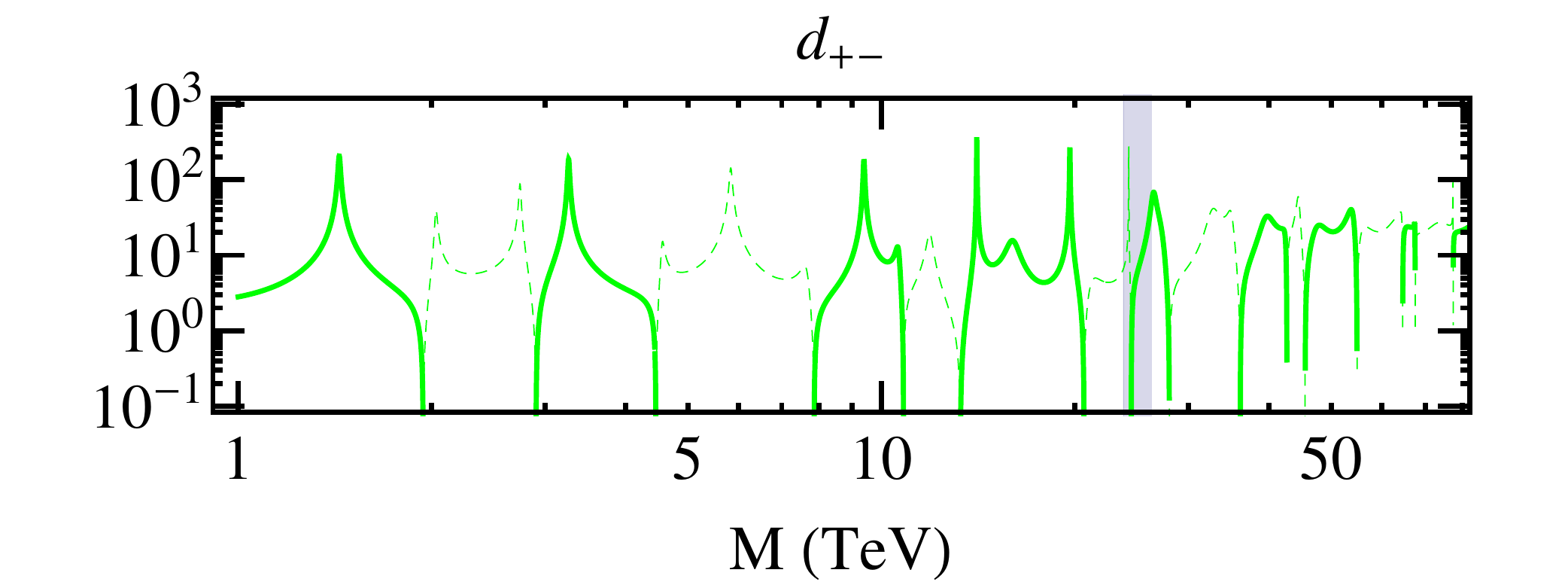}\\
\includegraphics[width=12.5cm]{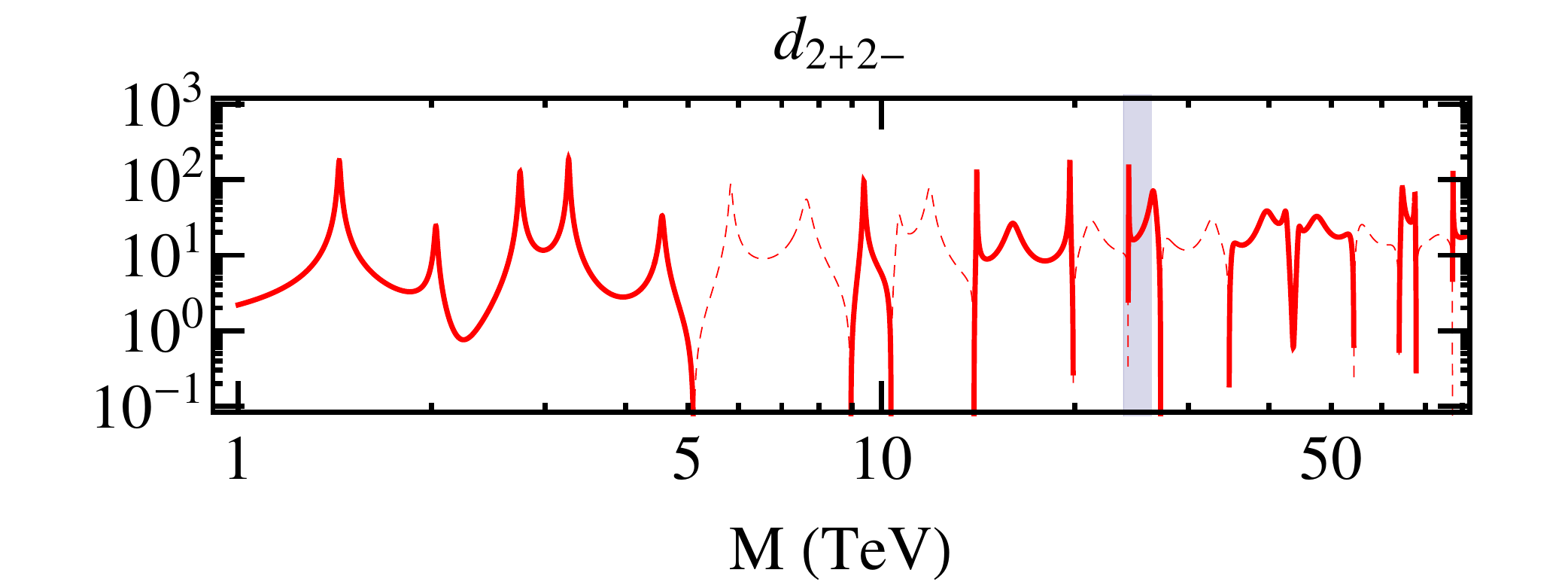}\\
\includegraphics[width=12.5cm]{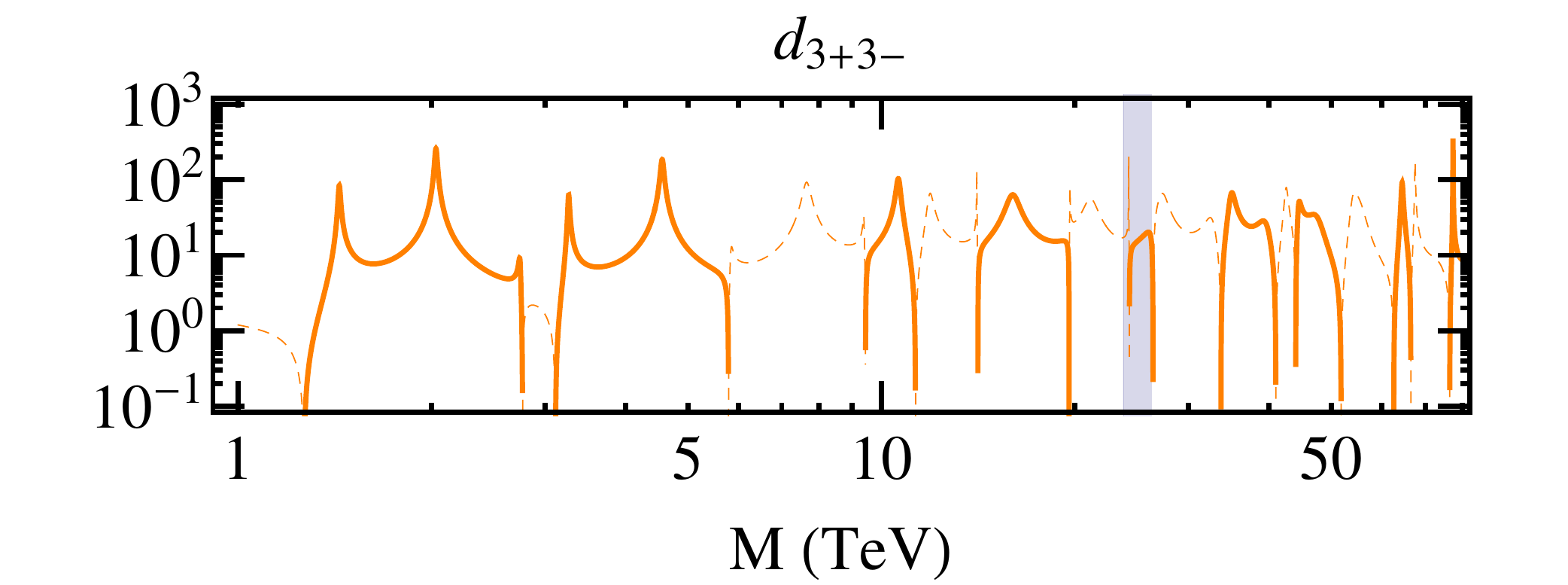}
\caption{Same as Fig.~\ref{fig:Qd}, but for the scalar 7-plet.}
\label{fig:Ed}
\end{center}
\end{figure}

We describe in Appendix \ref{sec:appA} a novel procedure for solving Eq.~\eqref{SoDE} which cures the numerical instabilities which plague the calculations when $\Delta /M \ll 1 $ \cite{Cohen:2013ama}. Using this method, we calculate the Sommerfeld enhancement factors for the 5-plet (7-plet) for masses between $1$ TeV and 30 TeV (75 TeV),  which approximately corresponds to three times the mass required by thermal production. The resulting values are  shown  in Figs.~\ref{fig:Qd} and \ref{fig:Ed}. The solid (dashed line) indicates the values of the mass where the enhancement factors are positive (negative), while the gray band indicates the value of the mass reported in Ref.~\cite{Cirelli:2007xd} that leads to the observed DM abundance via thermal freeze-out, and which approximately corresponds to $M\approx 10$ TeV for the 5-plet and $M\approx 25$ TeV for the 7-plet. As apparent from the plot, and as  expected for the non-perturbative exchange of light particles, we find multiple resonances across the TeV scale in both cases.

%%%%
\section{Annihilation Cross Sections }
\label{sec:cross}

We consider now the cross sections that are relevant for calculating the gamma-ray spectrum from DM annihilations in the Galactic Center. The prompt gamma-ray flux receives three main contributions: {\it i)}  a soft featureless part consisting of secondary photons from the decay and fragmentation of the $W$ and $Z$ pairs that are produced in DM annihilations, {\it ii)} the mono-energetic photons produced in the processes $\DM \,\DM\to\gamma\gamma$ and $\DM\, \DM \to\gamma Z$, and {\it iii)} the photons produced in the internal bremsstrahlung process $\DM\, \DM \to W^+W^-\gamma$.  The last two contributions are particularly relevant for indirect dark matter detection since they produce sharp spectral features in the gamma-ray spectrum for energies close to the DM mass, which might stand out over the featureless astrophysical background, thus offering a powerful probe of dark matter models~\cite{Srednicki:1985sf,Rudaz:1986db,Bergstrom:1988fp,Bergstrom:1989jr,Flores:1989ru,Bringmann:2007nk,Ibarra:2012dw}.

The s-wave cross section for DM annihilation into a two-body final state reads
\begin{equation}
(\sigma v)_{V_1 V_2}  = 2 \,d^*\, \Gamma_{V_1 V_2}\, d^T\,,\hspace{20pt} \text{with }\hspace{20pt} \Gamma_{V_1 V_2} = \frac{\xi}{32\pi M^2} \, {\cal W}_{V_1\,V_2}^\dagger  {\cal W}_{V_1\,V_2} \,,  
\label{SEsigmav}
\end{equation}
where the vector of Sommerfeld enhancement factors $d$ was calculated in Section \ref{sec:SE} and ${\cal W}_{V_1\,V_2}$ are  row vectors given in Table \ref{table:GandV}, which follow from the tree-level annihilation amplitude of the different pairs into a given gauge boson pair $V_1 \, V_2$.   Here, $\xi=1$ for Majorana DM whereas  $\xi=2$ for scalar DM. 

\begin{figure}[h!]
\begin{center}
\includegraphics[width=15.0cm]{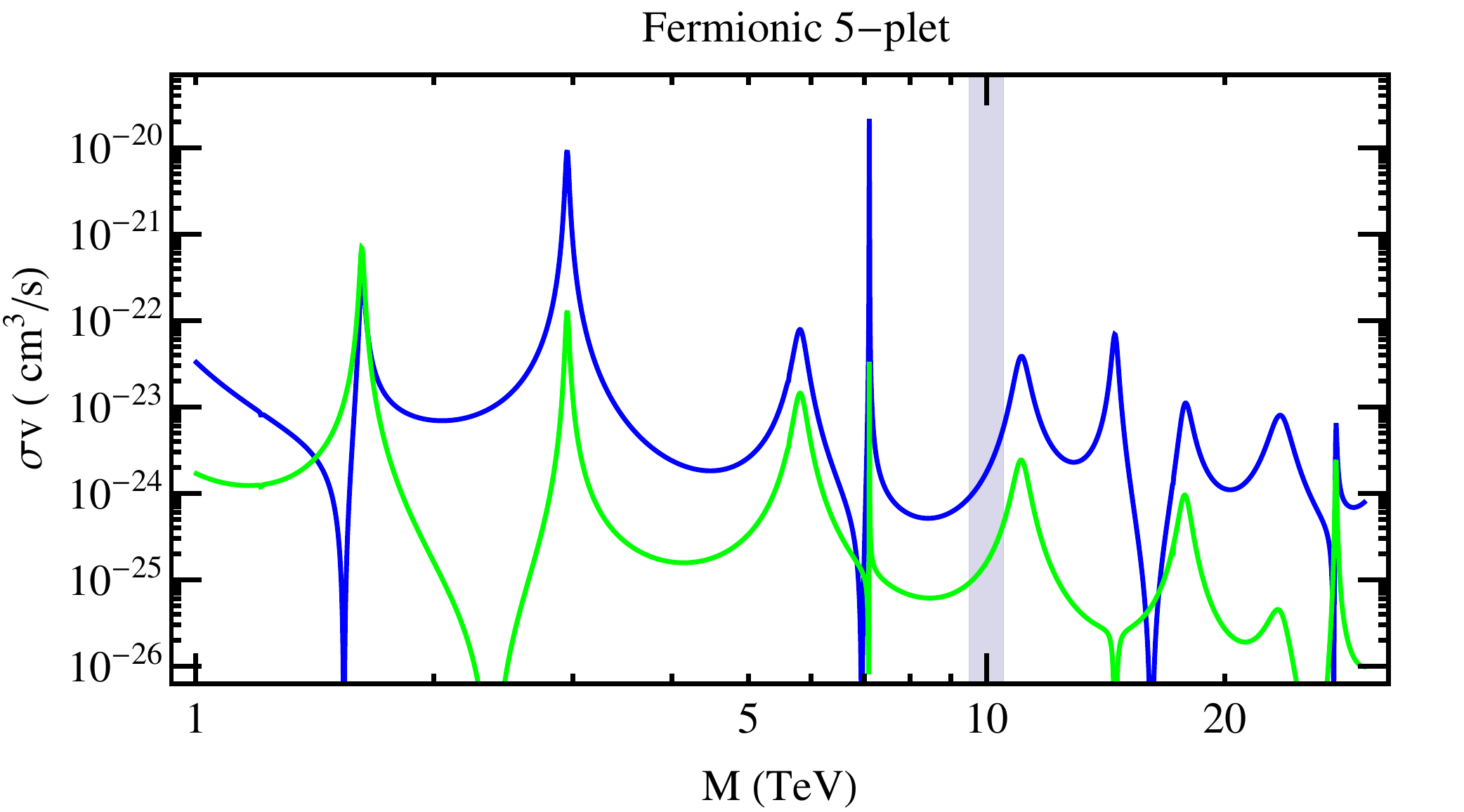}\\
\includegraphics[width=15.0cm]{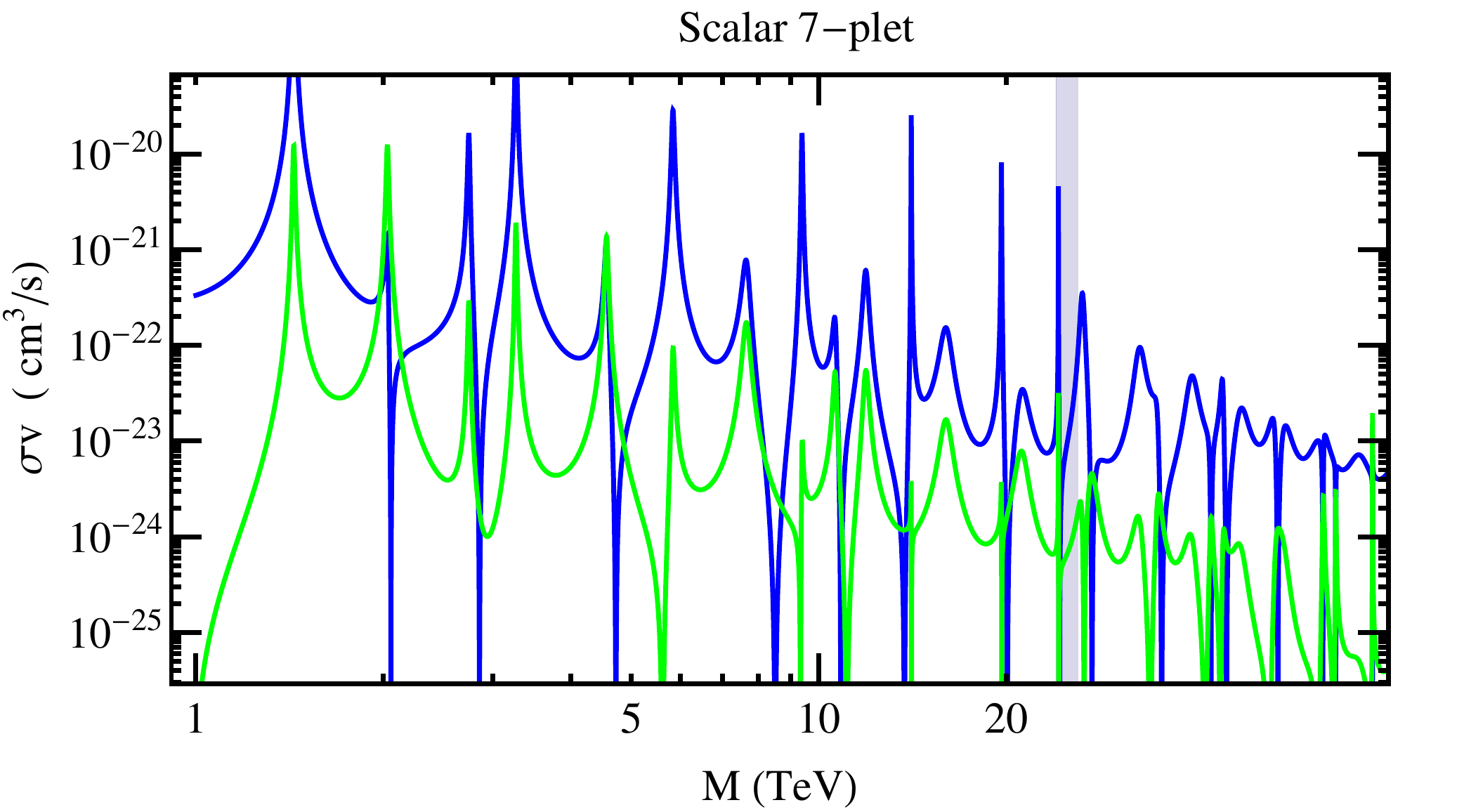}
\caption{DM annihilation cross section into $W^+W^-$ (blue line) and $\gamma\gamma$ (green line) for the fermionic 5-plet (upper panel) and the scalar 7-plet (lower panel).
}
\label{fig:Cross}
\end{center}
\end{figure}

We show in Fig.~\ref{fig:Cross} the DM annihilation cross sections into $W^+W^-$ and $\gamma\gamma$ as a function of the DM mass for the fermionic 5-plet and the scalar 7-plet; the cross sections into $ZZ$ and $\gamma Z$ can be readily obtained from  $(\sigma v)_{\gamma\gamma}$ from the relations
\begin{equation}
(\sigma v)_{ZZ} = \frac{ (\sigma v) _{\gamma\gamma}}{\tan^4\theta_W}\,,
\hspace{30pt}
(\sigma v)_{\gamma Z} = \frac{ 2 (\sigma v) _{\gamma\gamma}}{\tan^2\theta_W}\,.
\label{coreq}
\end{equation}
The resonance structure at the TeV scale, originated by the non-perturbative effects, is apparent in Fig.~\ref{fig:Cross}. Furthermore, we find dips in the two-body annihilation channels, namely masses for which the cross sections vanish, and which result from the destructive interference between Sommerfeld factors (this behavior, known as Ramsauer-Townsend effect, was also noted in  \cite{Chun:2012yt}). For instance, the 5-plet cross section for the $\gamma\gamma$ channel reads
\begin{equation}
(\sigma v)_{\gamma\gamma}  = \frac{2\pi\xi \alpha^2}{M^2} \, \left|d_{+-}\,+\,4\,d_{2+2-} \right|^2 \,.
\label{SEsigmavex}
\end{equation}
Therefore, for the mass ranges where the Sommerfeld factors have opposite signs, the cancellation between the two terms leads to a reduction of the cross section. An analogous effect also  occurs  for other annihilation channels. These cancellations are particularly significant for the scalar 7-plet, which presents dips in the $W^+ W^-$ and $\gamma\gamma$ cross sections for values of the DM mass comparable to the thermal value; we have checked numerically that the position of the dips does not change significantly when the relative velocity of the  DM particles is increased or reduced. On the other hand, for the fermionic 5-plet we find no cancellations in the cross sections in the mass range of relevance for thermal production. 

In addition to the two-to-two annihilation channels, in this class of models two-to-three annihilations can play an important role in indirect searches. Indeed, and due to the mass degeneracy among the components of the DM multiplet, the internal bremsstrahlung generically leads to a bump in the gamma-ray energy spectrum close to its kinematical endpoint, which might stand out over the astrophysical background. The Sommerfeld enhanced internal bremsstrahlung cross section reads
\begin{eqnarray}
&\displaystyle{\frac{d(\sigma v)_{WW\gamma}} {dE_\gamma}}
& = \frac{\xi}{128\pi^3 M^2} \int^{E_{W^+}^\text{max}}_{E_{W^+}^\text{min}} \left| 
\frac{1}{\sqrt{2}}d_{00}\,{\cal M} \left(\DM\DM \to  W^+W^-\gamma\right)\right.\nonumber\\
&&+\left.\sum_i d_{i+\,i-}{\cal M} \left(\DM^{i+} \DM^{i-} \to  W^+W^-\gamma\right) \right|^2 dE_{W^+} \,.
\label{CrossIBDiff}
\end{eqnarray}
where $E_{W^+}^\text{min}, E_{W^+}^\text{max}$ are the minimum and maximum energies kinematically accessible to the $W$ boson produced in the annihilation, while $\cal M$ stands for the tree-level annihilation amplitude, which we obtain using FeynArts~\cite{Hahn:2000kx} with an implementation of the fermionic 5-plet and scalar 7-plet DM models  made with FeynRules~\cite{Christensen:2008py}.

In Fig.~\ref{fig:CrossDiff} we show, for illustration, the main components of the prompt gamma-ray spectrum produced in the annihilations of a fermionic 5-plet with $M=10$ TeV (left plot) and   scalar 7-plet with $M=25$ TeV (right plot), assuming an energy resolution of 10\%.

We also remark that for small  DM relative velocity, as is the case of annihilations at the Galactic Center, the Sommerfeld enhancement factors are identical for fermionic and for scalar DM, since the potential matrix entering the Schr\"ondiger equation for the relevant two-body states, namely those with total charge $Q=0$ and total spin $S=0$, is the same for a given $SU(2)_L$ representation. Therefore, for a given DM multiplet, the annihilation cross section for scalar DM is simply a factor of two larger than for Majorana DM, as a result of the different value of the parameter $\xi$ in Eqs.~(\ref{SEsigmav}),~(\ref{CrossIBDiff}). The differential annihilation cross section for the scalar 5-plet (fermionic 7-plet) DM candidate, as well as their experimental limits, can then be straightforwardly derived from the analysis presented in this paper for the fermionic 5-plet (scalar 7-plet).

%%%%%%%%
\newpage
\section{Current Limits from H.E.S.S.}
\label{sec:HESS}

The H.E.S.S. instrument has measured the gamma-ray flux in the direction of the Galactic Center in an energy range from $\unit[300]{GeV}$ to $\unit[30]{TeV}$, finding an excellent agreement with the background expectations, thus allowing to set constraints on the rate of  DM annihilations. The expected gamma-ray flux from annihilations in the solid angle $\Delta \Omega$ centered in given direction of the sky is given by
\begin{equation}
 \frac{d\phi_\gamma}{d E_\gamma}=\frac{1}{8 \pi M^2}\frac{d (\sigma v)  }{d E_\gamma}  J_\text{ann}\;,
\end{equation}
where  $J_\text{ann}$ is the astrophysical $J$-factor, calculated by integrating the square of the DM density profile of the Milky Way $\rho(\vec r)$ over the line of sight and the region of interest $\Delta \Omega$
\begin{equation}
 J_\text{ann}=\int_{\Delta \Omega} d \Omega \int_\text{los}ds \rho^2\;.
\end{equation}

In our analysis we employ the gamma-ray flux measured with the H.E.S.S. instrument in a target region consisting of a circle of $1^\circ$ radius centered in the Milky Way Center, excluding the Galactic Plane by requiring $|b|\geq 0.3^\circ$~\cite{Abramowski:2011hc, Abramowski:2013ax}. In order to bracket the astrophysical uncertainties in the calculation of the flux, we consider two different DM halo profiles: the cuspy Einasto profile \cite{Navarro:2003ew, Graham:2005xx} 
\begin{equation}
\rho_\text{Ein}=\rho_s \text{exp} \left\{ -\frac{2}{\alpha}\left[ \left(\frac{r}{r_s}\right)^\alpha -1 \right]\right\}\,,
\end{equation}
with scale radius $r_s=\unit[20]{kpc}$ and shape parameter $\alpha=0.17$, and the cored isothermal profile 
\begin{equation}
\rho_\text{Iso}=\frac{\rho_s}{1+(r/r_s)^2}\;,
\end{equation}
where $r_s=\unit[5]{kpc}$  \cite{1980ApJS...44...73B}. Here,  $\rho_s$ is chosen such that the  DM density at the position of the Sun, which we consider located at a distance  $d_\odot=\unit[8.5]{kpc}$ from the Galactic Center, reproduces the local DM density $\rho_0=\unit[0.39]{GeV/cm^3}$~\cite{Catena:2009mf}. Then, the $J$-factors are $J_\text{Ein}=4.43\times 10^{21}$GeV$^2$\,sr\,$/$cm$^5$  \cite{Abramowski:2011hc} for the Einasto profile and $J_\text{Iso}=3.23\times 10^{19}$GeV$^2$\,sr\,$/$cm$^5$  for the isothermal profile. 

\begin{figure}[t]
\begin{center}
\includegraphics[trim=0cm 0cm 0cm 1.15cm,clip,
width=7.4cm]{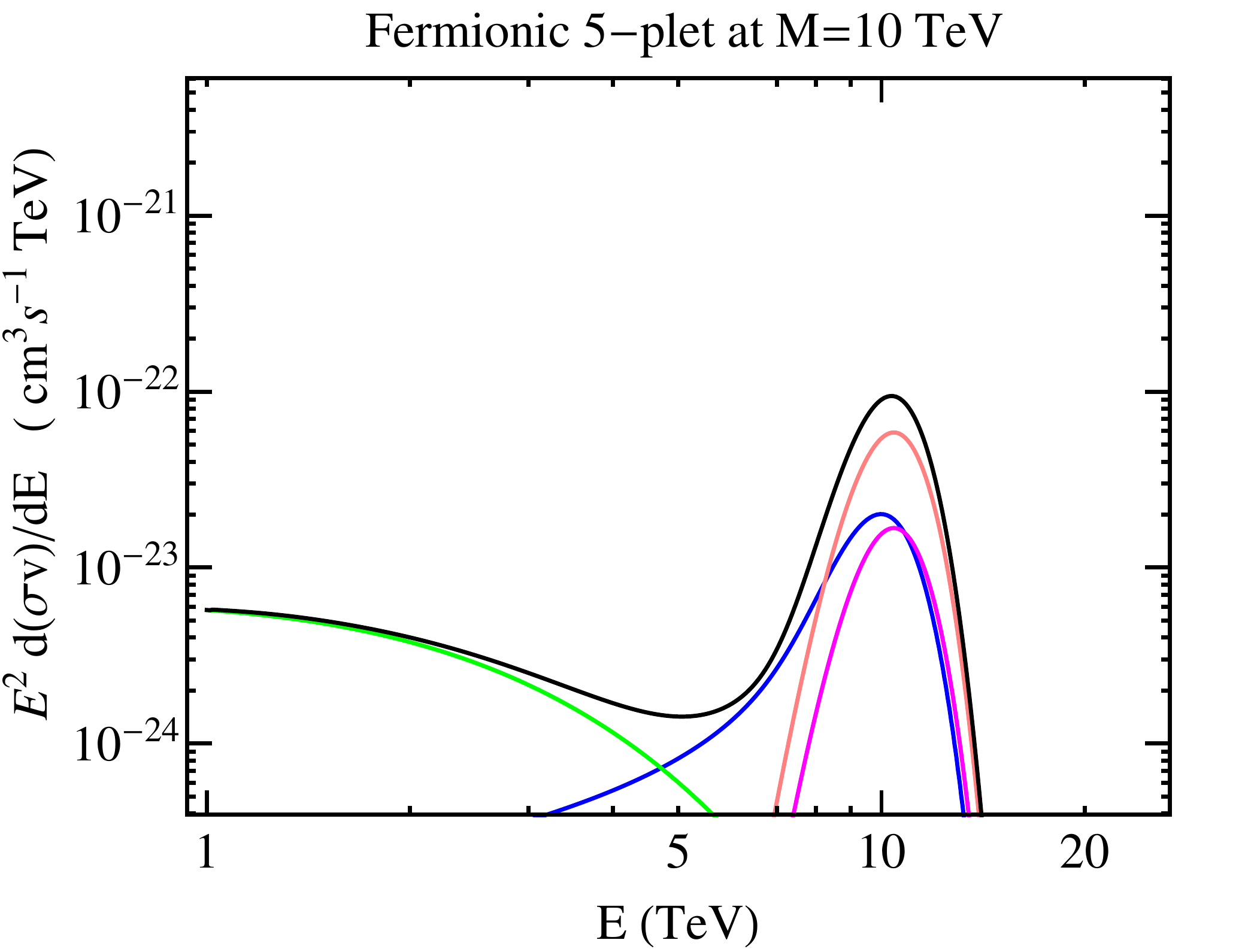}
\includegraphics[trim=0cm 0cm 0cm 1.15cm,clip,
width=7.4cm]{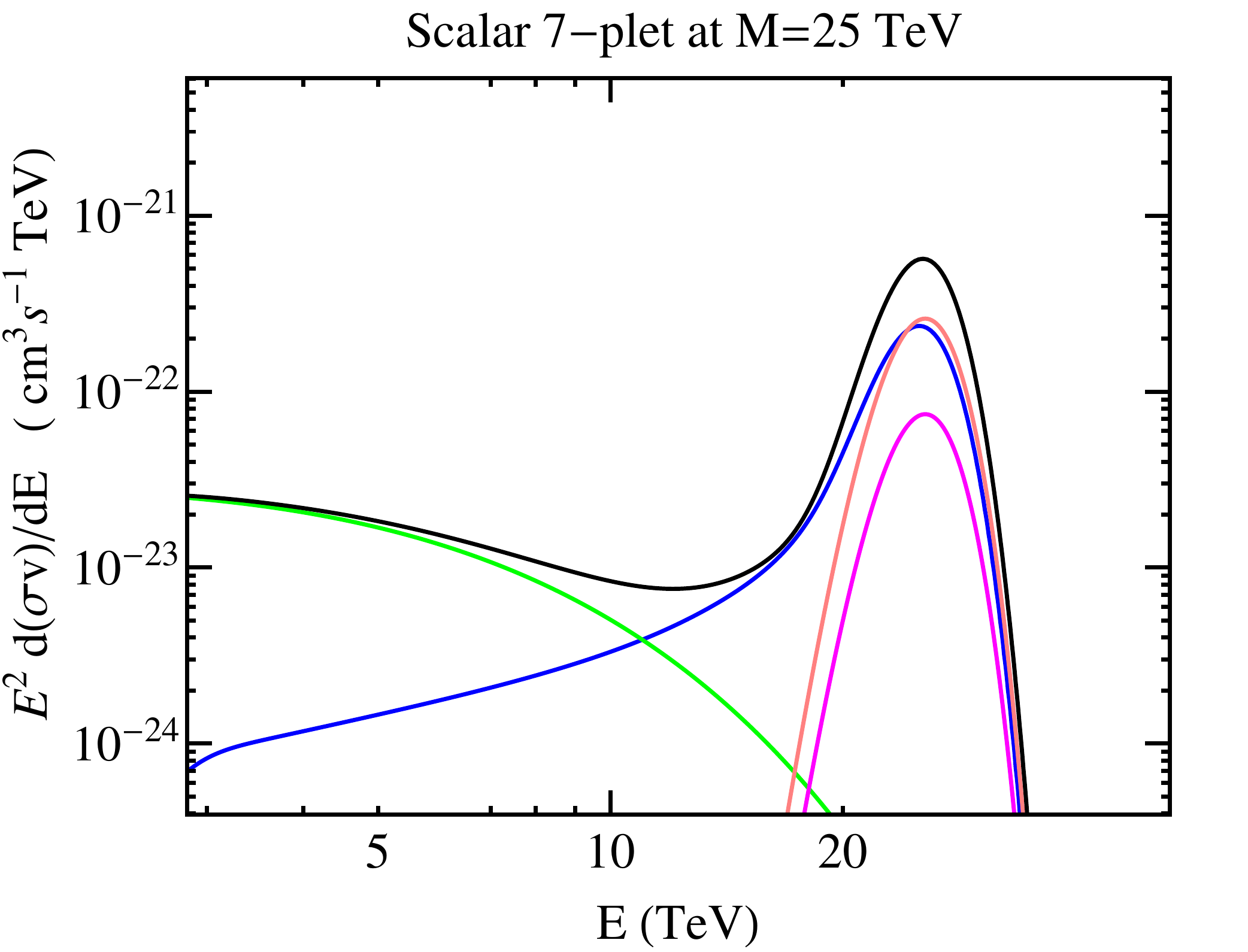}
\caption{Differential annihilation cross sections into $\gamma \gamma$ (magenta line), $\gamma Z$ (pink line),  $W^+W^-\gamma$ (blue line)  and continuum from $ZZ$ and $W^+W^-$ (green line), as well as the total differential cross section (black line), for a fermionic 5-plet DM candidate with mass 10 TeV (left panel) and a scalar 7-plet with mass 25 TeV (right panel). We assume an energy resolution of 10\% which is  typical for current gamma-ray telescopes.
}
\label{fig:CrossDiff}
\end{center}
\end{figure}

On the other hand, the differential cross section to gamma-rays  $d(\sigma v) _\gamma/dE_\gamma$  consists of three parts, as mentioned in the previous section. First,  a continuum of gamma-rays, which is produced in DM annihilations into $W$ and $Z$ bosons, and which is calculated from the product of the cross section into $W^+W^-$ or $ZZ$ and the differential number of photons produced per annihilation, which we take as $dN_\gamma/dx= 0.73\,x^{-1.5}\,e^{-7.8x}$, with $x=E_\gamma/M$~ \cite{Bergstrom:1997fj}. Second, the mono-energetic photons produced in annihilations into $\gamma Z$ and $\gamma\gamma$, which would be observed with H.E.S.S. as Gaussian distributions, normalized to 1 and 2 respectively, with a width given by the energy resolution of the instrument, which is approximately 17\% at $\unit[500]{GeV}$ improving to 11\% at $\unit[1]{TeV}$ \cite{Abramowski:2013ax}. Finally, the line-like contribution from internal bremsstrahlung, given by Eq.~\eqref{CrossIBDiff}, which we also smear out with the energy resolution of the instrument.  In Fig.~\ref{fig:CrossDiff} we show, for illustration, the differential cross sections for red a fermionic 5-plet with $M= \unit[10]{TeV}$ and a scalar 7-plet with $M= \unit[25]{TeV}$, which approximately correspond to the thermal masses of these two DM candidates. 

We now calculate constraints on  the fermionic 5-plet and scalar 7-plet  DM models which follow from the non-observation by  H.E.S.S.  of a sharp gamma-ray spectral feature nor an exotic featureless contribution to the gamma-ray flux. To calculate limits on the DM annihilation cross section into sharp spectral features, we adopt the phenomenological background model proposed by the H.E.S.S. collaboration~\cite{Abramowski:2013ax}, which is described by 7 parameters. We then include the sharp spectral feature from DM annihilation and calculate the  95\% C.L. one-sided limit on the cross section. On the other hand, we calculate constraints on the featureless component from $W^+W^-$ or $ZZ$ annihilations following  \cite{Abramowski:2011hc}, which compares the  gamma-ray fluxes measured with the H.E.S.S. instrument in a ``search region'' and in a ``background region''. The inferred residual flux is consistent with zero, thus allowing to  derive upper limits on the flux from annihilations. We note that this approach is only valid for cuspy profiles, since for cored profiles the signal gets completely subtracted, therefore in this case we will only present limits for the Einasto profile. 

\begin{figure}[h!]
\begin{center}
\includegraphics[width=15.5cm]{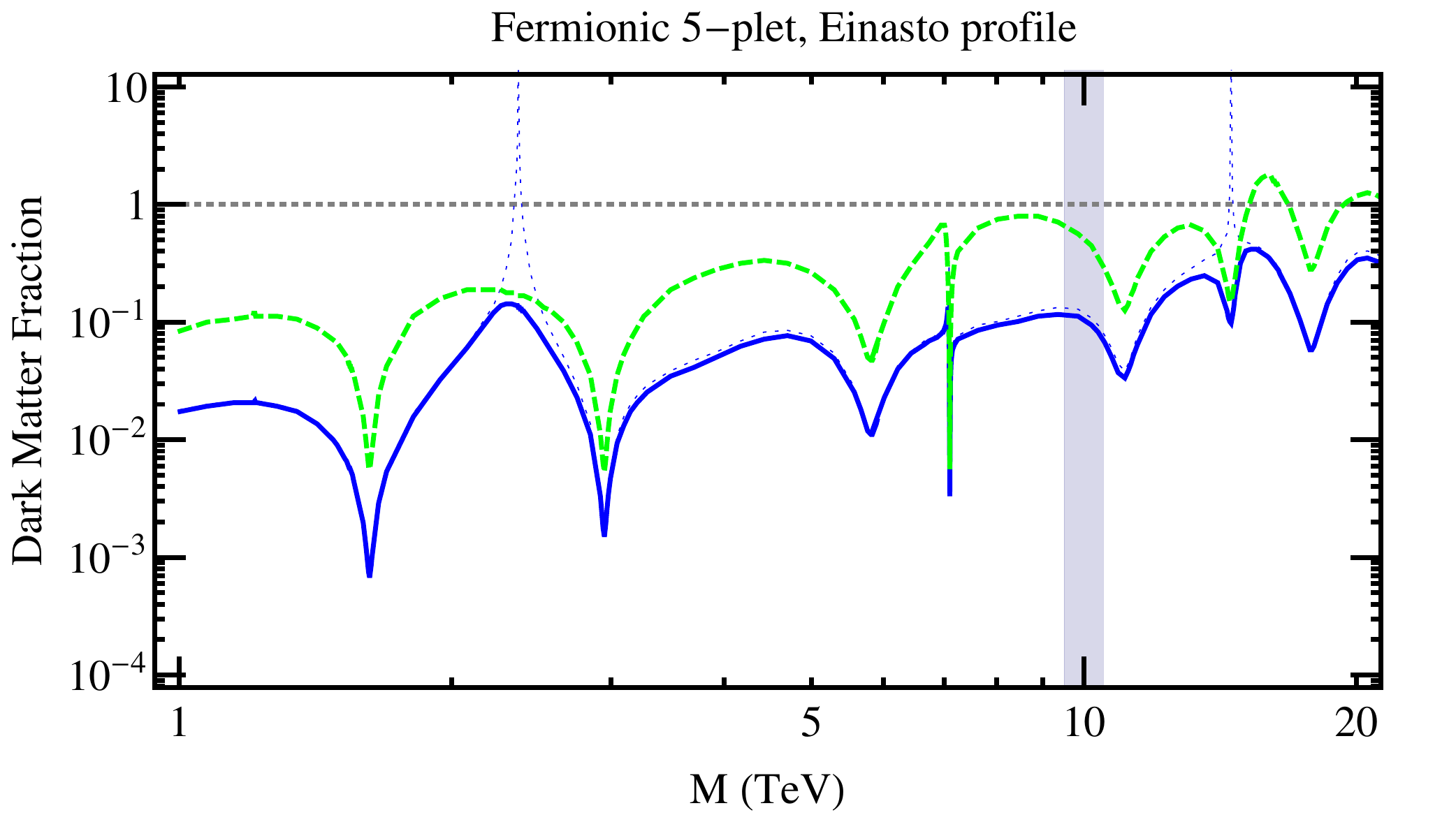}\\
\includegraphics[width=15.5cm]{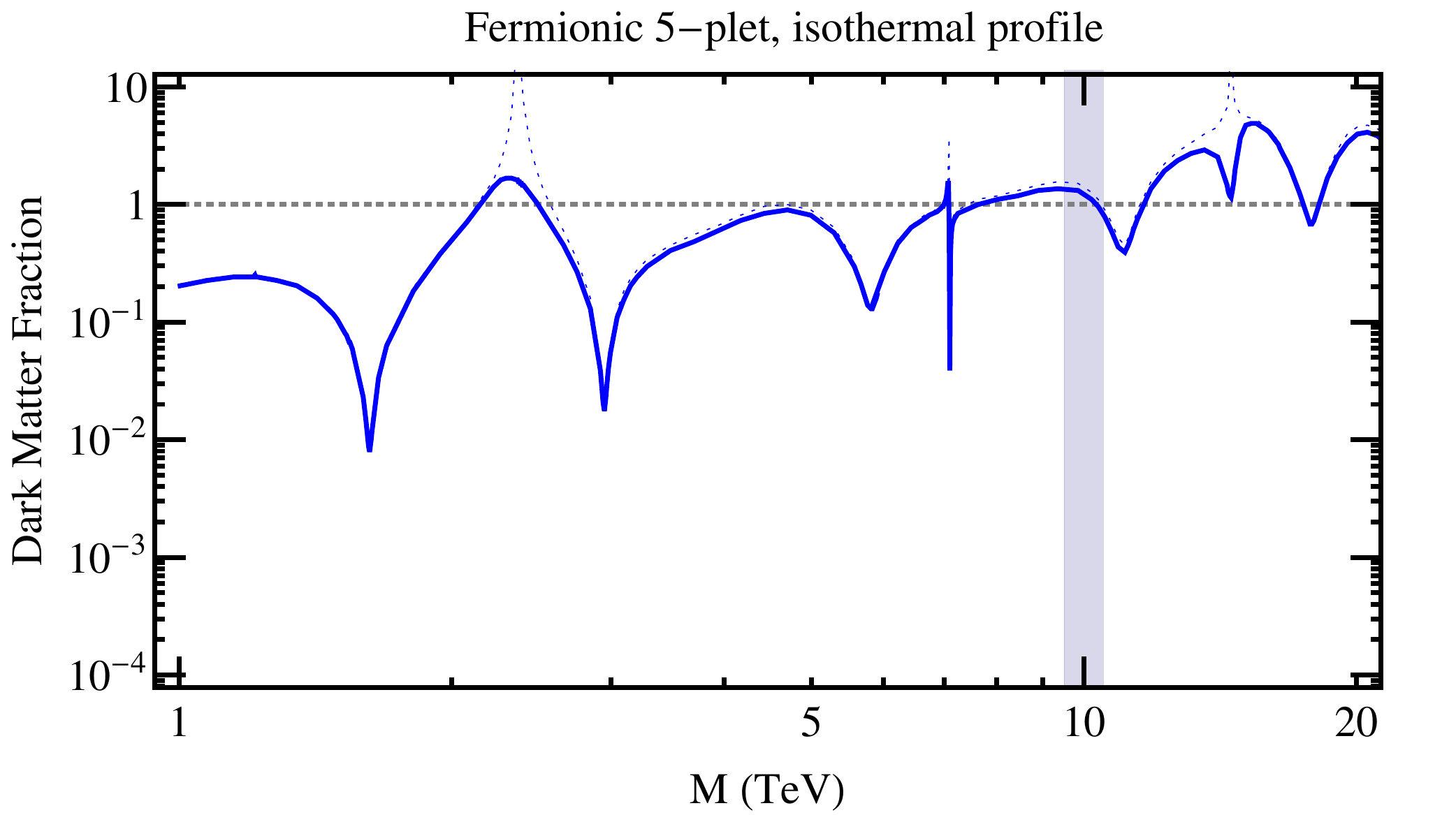}
\caption{\small 95\% C.L. limits on the dark matter fraction for the fermionic 5-plet from the non-observation by H.E.S.S. of sharp gamma ray spectral features, including (solid blue line) and neglecting (dotted blue line) the internal bremmstrahlung contribution, as well as from the non-observation of the continuum gamma-rays from annihilations into  $W^+W^+$ and $ZZ$ (green line), assuming the Einasto profile (upper panel) and the isothermal profile (lower panel). 
}
\label{fig:HESS5}
\end{center}
\end{figure}

\begin{figure}[h!]
\begin{center}
\includegraphics[width=15.5cm]{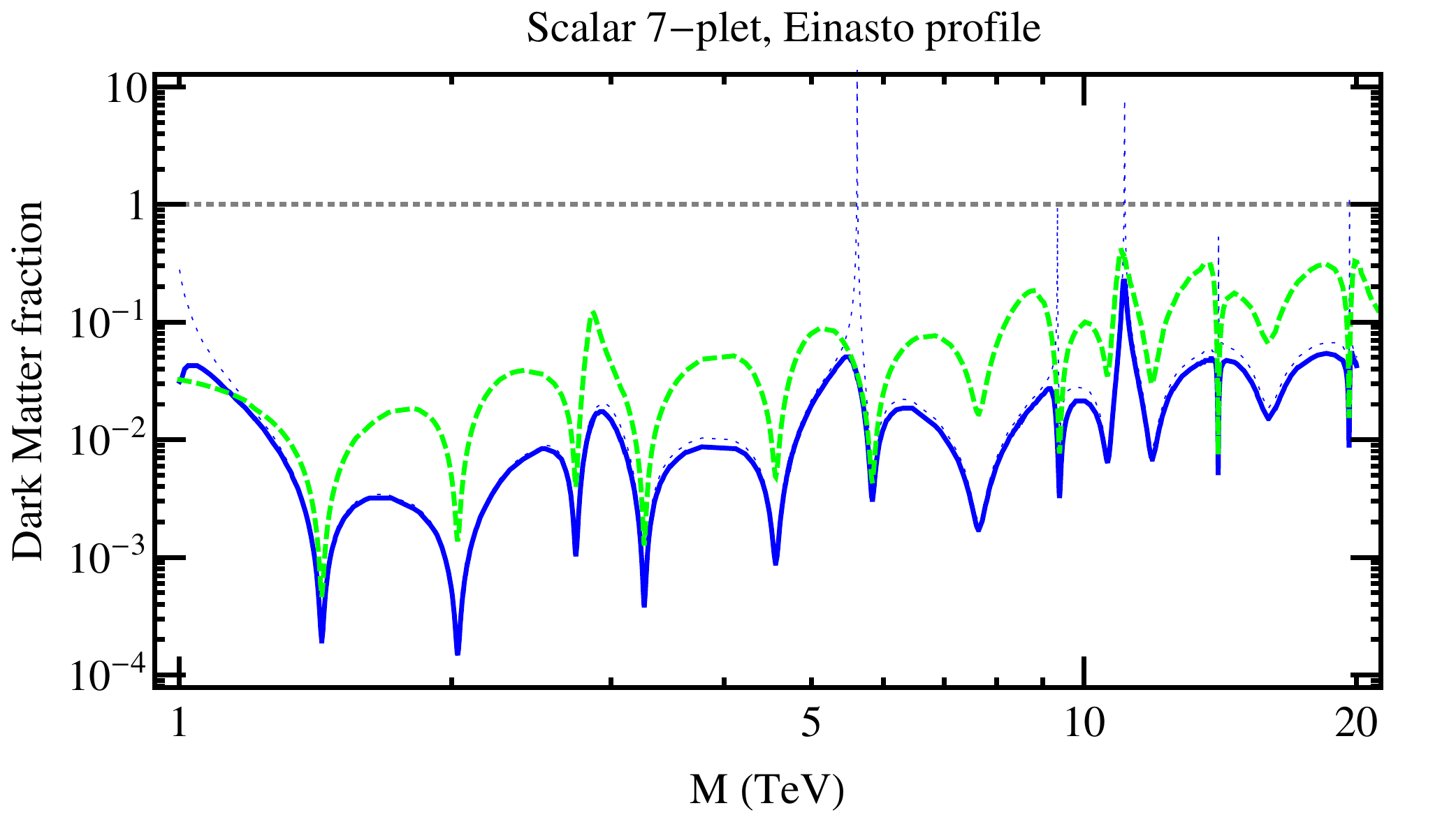}\\
\includegraphics[width=15.5cm]{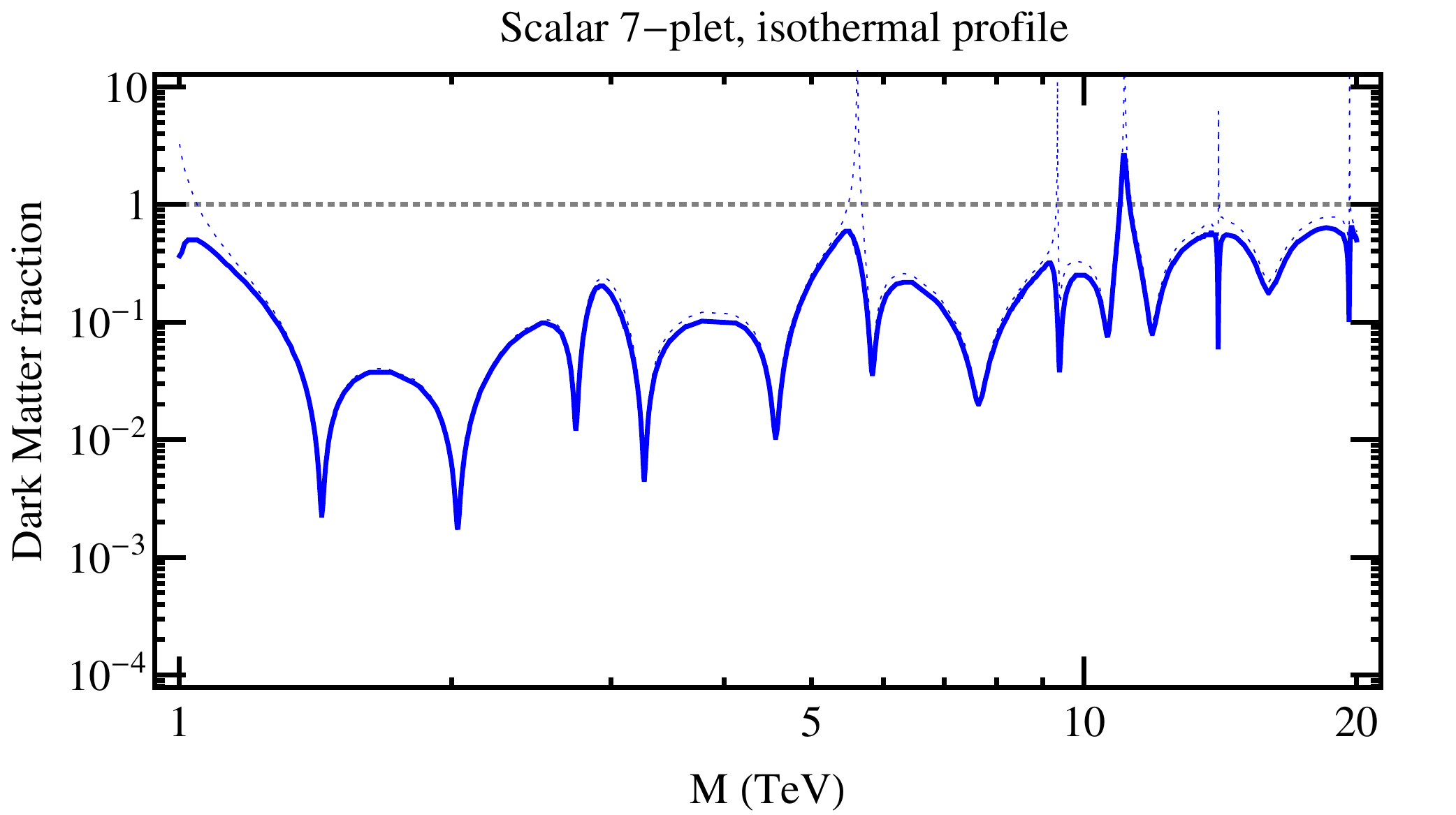}
\caption{\small Same as Fig.~\ref{fig:HESS5}, but for the scalar 7-plet.
}
\label{fig:HESS7}
\end{center}
\end{figure}

We show the impact of the H.E.S.S. measurements on  the fermionic 5-plet and scalar 7-plet  DM models in  Figs.~\ref{fig:HESS5} and \ref{fig:HESS7}, respectively, for the Einasto and isothermal profiles, and separating the limits from the non-observation of sharp spectral features (blue solid lines) or from an excess in the continuum (green solid lines). We present the constraints in terms of the DM fraction, which is given  by the square root of the signal normalization factor. We highlight the role of the internal bremsstrahlung in probing  the fermionic 5-plet and scalar 7-plet models by showing also the limit obtained when only the gamma-ray lines from annihilations into $\gamma \gamma$ and $\gamma Z$ are included (dotted blue line). As apparent from the plots, the limits from lines exhibit peaks, which correspond to the dips in the annihilation cross section shown in  Fig.~\ref{fig:Cross}, and where cancellations among amplitudes occur. On the other hand, the annihilations into $W^+W^-\gamma$ are not affected by these cancellations, thus allowing to probe DM masses which are otherwise unconstrained by line searches. Furthermore, after including the $W^+W^-\gamma$ channel, we find that for DM masses above $\sim 1.2$ TeV the limits from sharp spectral features are more stringent than those from the continuum,  concretely by a factor of $1.5\sim 8.5 $ for the fermionic 5-plet and $1.5\sim 11$ for the scalar 7-plet.

Our analysis excludes DM consisting exclusively of fermionic 5-plet or scalar 7-plet neutral states in the whole mass range between $M=1$ TeV up to $M=\unit[20]{TeV}$, assuming  the Einasto profile. More concretely, for low masses the limits are very stringent and exclude a DM fraction of $7 \times 10^{-4}$  ($1.5 \times 10^{-4}$) for the 5-plet (7-plet) at $M=\unit[1.6]{TeV}$ ($\unit[2]{TeV})$.  In the case of the isothermal profile, we find some mass ranges where the 5-plet can be the dominant component of DM, notably the regions around the 5-plet thermal mass. The 7-plet, on the other hand, is excluded as the dominant component of  DM for masses between 1 and 20 TeV, with the exception of a narrow mass window around $\unit[12]{TeV}$. Thermal masses for the 7-plet are, unfortunately, not accessible to the  H.E.S.S. instrument, although they may be probed by CTA, as we will discuss in the next section.

\newpage
\section{Prospects with CTA}
\label{sec:CTA}

We consider in this section the sensitivity to the fermionic 5-plet or scalar 7-plet DM candidates of the  upcoming Cherenkov Telescope Array, an instrument that will be able to measure the  gamma-ray flux up to energies $\sim\unit[100]{TeV}$. In our analysis we will employ the instrument properties reported in \cite{Bernlohr:2012we} for the array I, a balanced array that can measure gamma-rays in a wide energy range from a few tens of GeV up to over 100 TeV. Its effective area is larger than $\unit[10^6]{m^2}$ above $\unit[1]{TeV}$ with a resolution better than 10\% in this energy range.\footnote{The CTA collaboration has recently released an updated performance for a different array configuration~\cite{CTA}. We have checked that, with respect to array I, the corresponding limits on the DM fraction for gamma-ray monochromatic lines improve up to a factor of 2.8 for low masses and up to a factor of 1.3 above 4 TeV.}

Since in these scenarios the limits from continuum gamma-rays are  in general worse than the limits from sharp spectral features, we will only consider the latter for our CTA forecast. In our analysis we closely follow the approach described in \cite{Ibarra:2015tya}. First, we generate mock data. The expected number of counts in the energy bin $i$ is given by
\begin{equation}
\label{eq:Counts}
 n_i=\Delta t \int_{\Delta E_i}dE \int dE' R(E,E') A_\text{eff}(E')\frac{d\Phi_\text{tot}}{d E_\gamma}, 
\end{equation}
where $\Delta t$ is the observation time in the region of interest, $\Delta E_i$ the width of the $i$th energy bin and $\Phi_\text{tot}$ the total background in the region of interest. The energy resolution $R(E,E')$ and effective area $A_\text{eff}$ of the instrument are taken from \cite{Bernlohr:2012we}. 
In this analysis we use $200$ energy bins per decade in order to resolve the sharp spectral features. Then, from the mean number of counts in each energy bin, we generate 200 sets of mock data using Poisson distributed random numbers. 

In order to compare the CTA reach with the H.E.S.S. constraints, we consider the same region of interest around the Galactic Center and 112h of observation time. For the background we make use of the parametrization presented in \cite{Ibarra:2015tya}, which takes into account not only the diffuse gamma-rays and the H.E.S.S. point source in the Galactic Center, but also cosmic ray backgrounds. In fact, protons are so abundant in cosmic rays that they constitute an important contribution, especially at higher energies, even with good proton rejection efficiencies of $10^{-2}-0.2$. Electrons and positrons produce air showers that are indistinguishable from gamma-ray initiated showers, hence they comprise an irreducible background for Cherenkov telescopes. For our region of interest, electrons are the dominant background at low energies.

The limits are calculated using the sliding energy window technique. This method is based on the assumption that the background is locally well described by a power law. The energy range for which this is the case determines the size of the energy window. In our analysis we use the largest possible window size of $\epsilon=2$ \cite{Ibarra:2015tya}, which is suitable for the CTA background. Finally, inside these energy windows, we determine the 95\% C.L. one-sided upper limits using a profile likelihood analysis.

\begin{figure}[h!]
\begin{center}
\includegraphics[width=15.5cm]{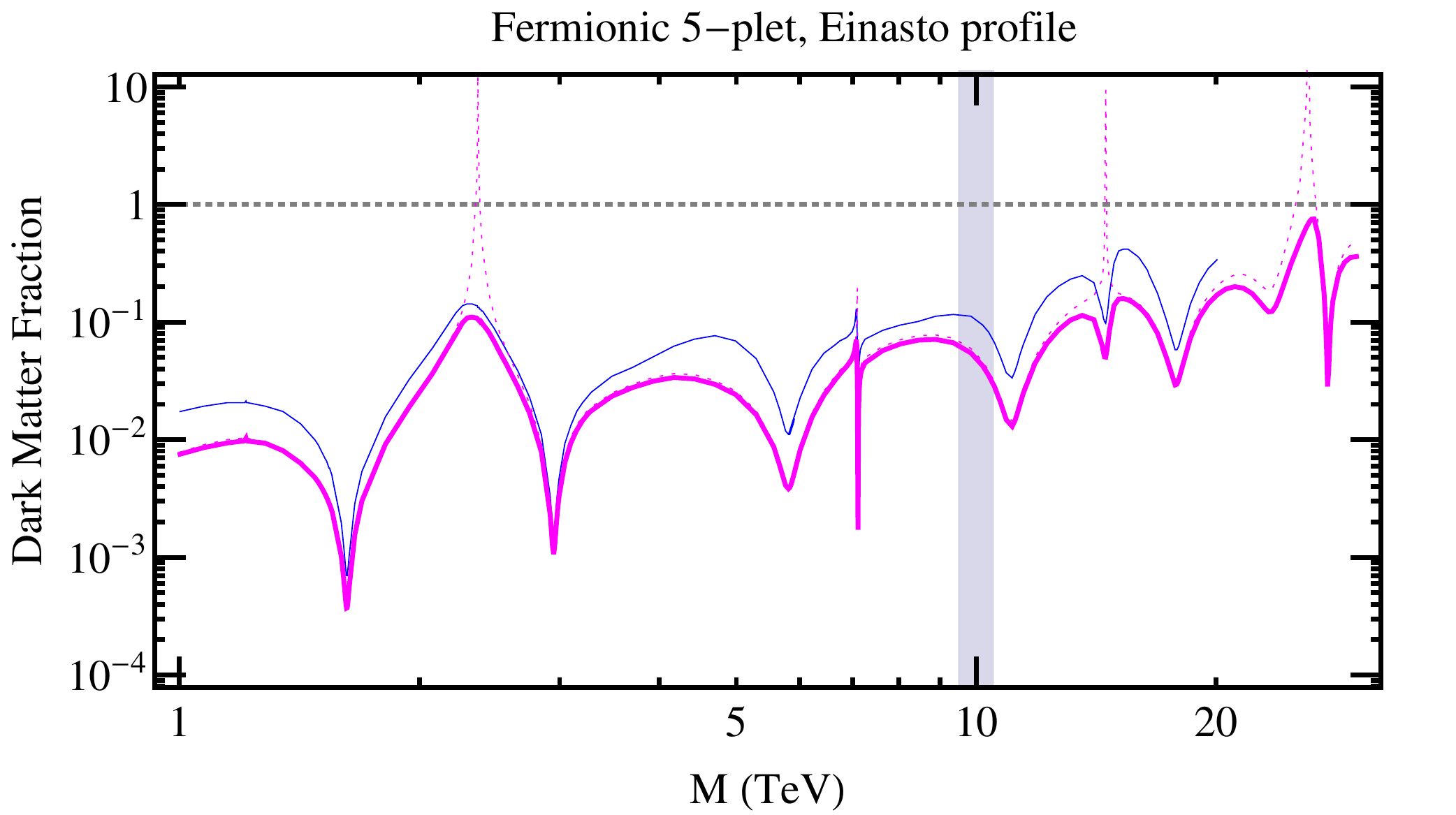}\\
\includegraphics[width=15.5cm]{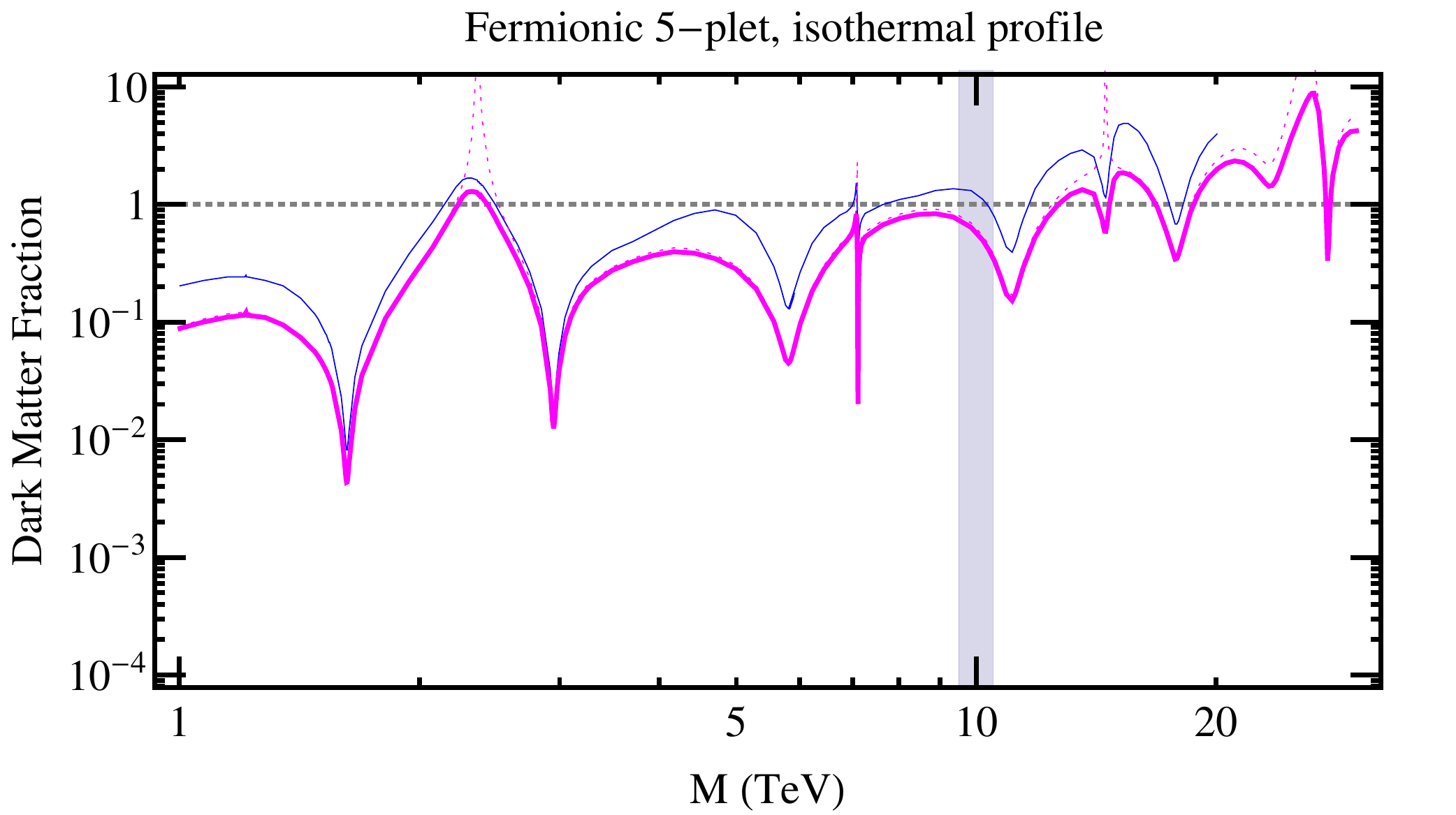}
\caption{\small 95\% C.L. projected limits on the dark matter fraction for the fermionic 5-plet from the non-observation by the upcoming CTA of a sharp spectral feature in the Galactic Center, including (solid magenta line) and neglecting (dotted magenta line) the internal bremmstrahlung contribution, assuming the Einasto profile (upper panel) and the isothermal profile (lower panel). We also show for comparison the current H.E.S.S. limits (solid blue line).
}
\label{fig:CTA5}
\end{center}
\end{figure}

\begin{figure}[h!]
\begin{center}
\includegraphics[width=15.5cm]{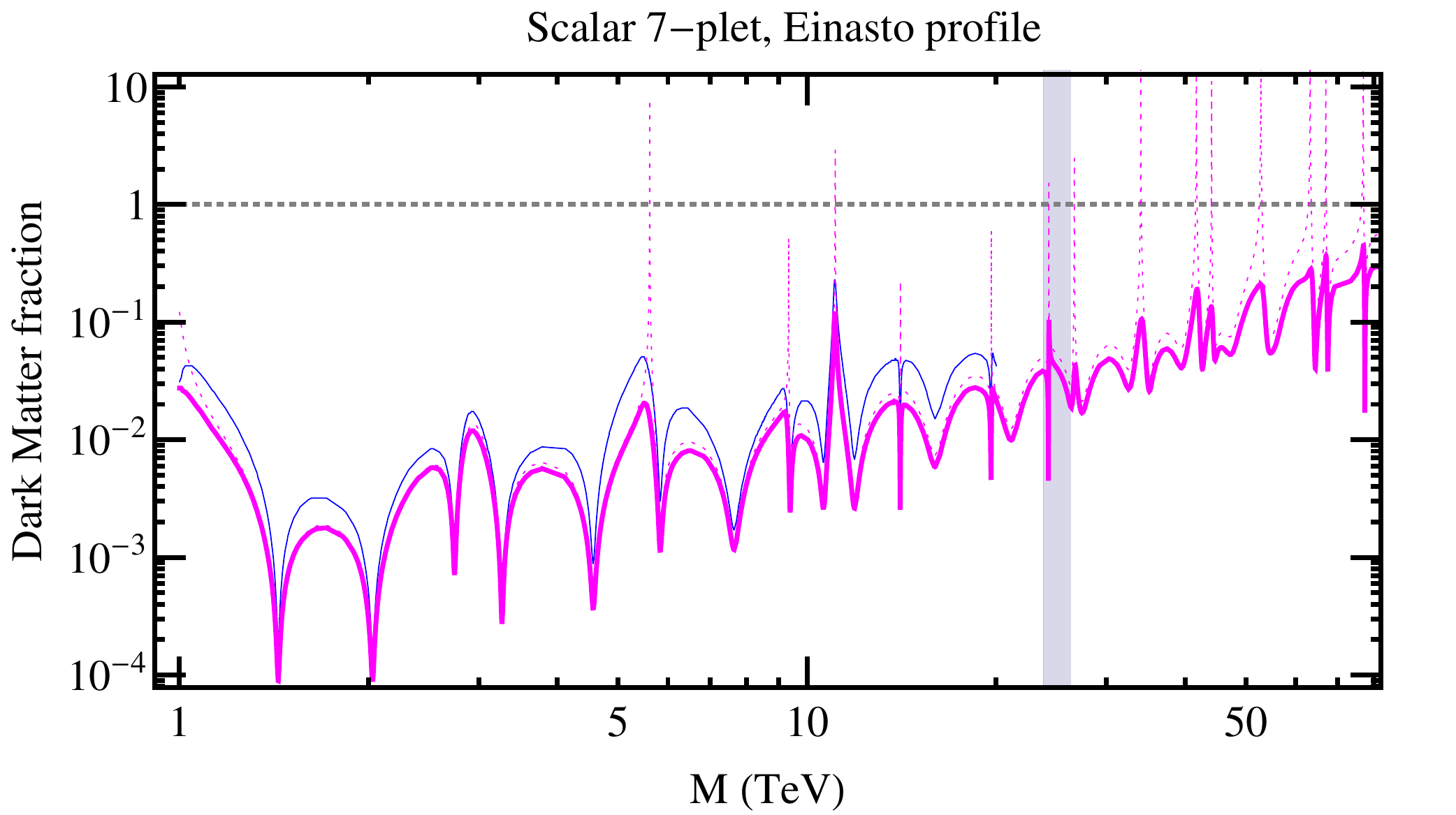}\\
\includegraphics[width=15.5cm]{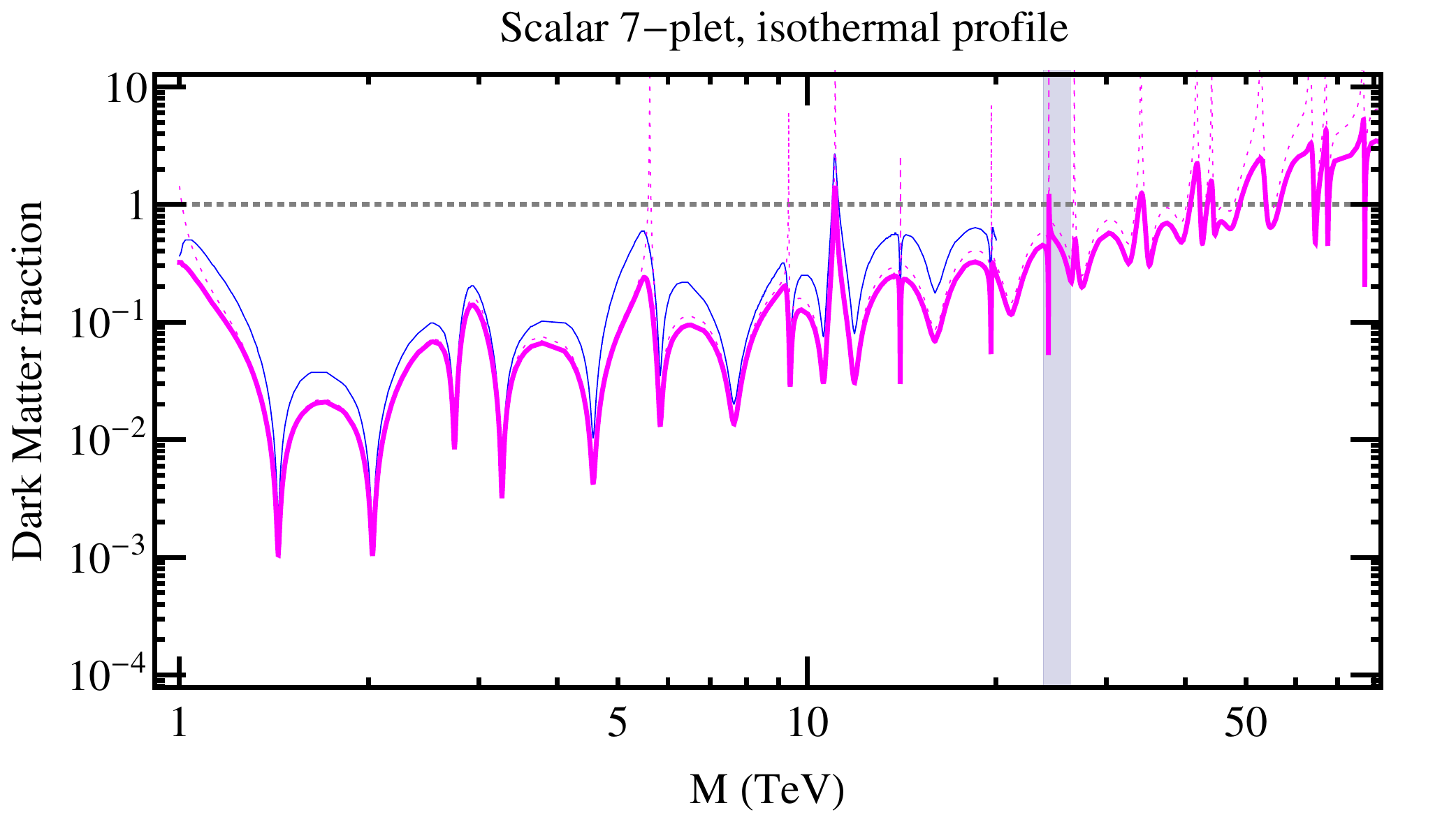}
\caption{\small Same as Fig.~\ref{fig:CTA5}, but for the scalar 7-plet. 
}
\label{fig:CTA7}
\end{center}
\end{figure}

The resulting limits for both profiles under consideration are shown  in Figs.~\ref{fig:CTA5} and \ref{fig:CTA7} for the fermionic 5-plet and scalar 7-plet, respectively. Again, we present the limits in terms of the DM fraction, which is obtained from logarithmically averaging the signal normalization factor in the 200 mock data realizations, and then taking its square root. We find that CTA can improve the H.E.S.S. limits on the DM fraction by a factor of $1.2\sim 3$ and close in on  scenarios with heavier masses. More specifically, even for the isothermal profile,  CTA will be able to  exclude 5-plet masses up to $\unit[13]{TeV}$ and 7-plet masses up to $\unit[40]{TeV}$. In particular, CTA will be able to exclude the possibility that the 7-plet is produced via thermal freeze-out. On the other hand, for the Einasto profile, DM entirely consisting of 5-plet and 7-plet states can be excluded up to 30 TeV and 75 TeV, respectively, which roughly corresponds to three times the values of the mass favored by thermal production.

\clearpage
%%%%%%%%%%%%%
\section{Conclusions}
\label{sec:conclusions}

In this work we have studied the gamma-ray spectrum generated in the annihilation  of fermionic 5-plet or scalar 7-plet dark matter particles, motivated by the Minimal Dark Matter framework.
We have remained agnostic about the dark matter production mechanism in the Early Universe, hence the dark matter mass is a free parameter in our analysis, which we take between 1 TeV and 30 TeV (75 TeV) for the 5-plet (7-plet), which roughly corresponds to three times the mass expected for a thermal relic. In this mass range, non-perturbative effects are expected to be significant and ought to be taken into account. In this work, we have also developed a new algorithm to calculate the Sommerfeld enhancement factors in the limit of low relative velocities, which cures the numerical instabilities reported in previous approaches. 

We have included in our analysis the internal bremsstrahlung, which produces a sharp spectral feature in the gamma-ray spectrum as a natural consequence of the small mass splitting between the charged and neutral states of the dark matter multiplet. This component leads to new opportunities to detect sharp gamma-ray spectral features in this class of  scenarios, since the cross sections into $\gamma\gamma$ and $\gamma Z$ vanish for certain values of the DM mass, due to cancellations among the amplitudes contributing to these channels.

We have then confronted the fermionic 5-plet and scalar 7-plet dark matter scenarios with  the most recent Galactic Center observations of the H.E.S.S. instrument, considering the continuum of gamma-rays from annihilations into $W^+W^-$ and $ZZ$, as well as the sharp spectral features from monochromatic photons and internal bremsstrahlung.  Assuming the Einasto profile, the H.E.S.S. measurements already exclude  DM consisting exclusively  of fermionic 5-plet or scalar 7-plet neutral states in the mass range from $1$ TeV up to $20$ TeV, which includes in particular the mass expected for a thermally produced 5-plet DM. We have also discussed the sensitivity of the upcoming  Cerenkov Telescope Array to this class of scenarios, in particular, the prospects to detect sharp gamma-ray spectral features. We find that CTA can improve the H.E.S.S. reach to the DM fraction in this class of models  by a factor of $1.2\sim 3$. For the Einasto profile, dark matter entirely consisting of 5-plet (7-plet) states can be probed in the mass range between 1 TeV and 30 TeV (75 TeV), and will be able to test these two scenarios assuming thermal DM production. For the isothermal profile, the prospects are poorer but can nonetheless probe large regions of the parameter space where either the 5-plet or the 7-plet constitutes the dominant component of dark matter.

We conclude highlighting the role of Cherenkov telescopes in probing scenarios with heavy dark matter particles, as is the case of scenarios motivated by the Minimal Dark Matter framework, and which are challenging to probe with direct detection experiments or with collider experiments.

%%%%%%%
\appendix

\section{Numerical Algorithm for Calculating the Sommerfeld Effect Today }
\label{sec:appA}

The Sommerfeld enhancement factors are obtained solving Eq.~(\ref{SoDE}) with the boundary conditions given in Eq.~(\ref{BC}). Although this is in principle straightforward, the numerical calculation of the solution is challenging, specially when $\Delta / M \ll 1$, due to the particular form of the boundary conditions~\cite{Cohen:2013ama}. Here we introduce a novel algorithm that cures the numerical instabilities that plague the calculation of the enhancement factors.

On the one hand, let us notice that Eq.~(\ref{SoDE}) can be rewritten as  
\begin{equation}
h'(r) +h(r)^2+ M \left(\dfrac{1}{4} M v^2 {1\!\!1} - V(r) \right)   = 0\,,\hspace{20pt}\text{where  }h(r) \equiv g'(r)g(r)^{-1}\,.
\label{SoDE2}
\end{equation}
with a boundary condition at large distances given by
\begin{equation}
h(\infty) = (i Mv/2)\sqrt{{1\!\!1}-4V(\infty)/(Mv^2)}.
\end{equation}
In this expression the eigenvalues are either purely imaginary or negative, so that the solution at infinity is either an out-going or a damping wave, respectively.  

On the other hand, conservation of probability on Eq.~(\ref{SoDE})  can be written as
\begin{equation}
i \frac{d}{dr}\left(g(r)^\dagger g'(r)- g(r)'^\dagger g(r) \right)=0\,,
\end{equation}
which along Eq. \eqref{BC} can be used to prove that 
\begin{equation}
d^\dagger\, d = \frac{1}{iMv}\left(h(0)-h(0)^\dagger\right)\,,
\label{ddeq}
\end{equation}
where  $d \equiv (\cdots \,\,d_{+-}\,\, d_{00})$.

Therefore, in order to calculate the enhancement factors,  instead of solving the second-order differential Eq.~\eqref{SoDE} with boundaries at two different points, it suffices to solve the first-order differential equation Eq.~\eqref{SoDE2} with one boundary at infinity.  With this solution, we calculate the right-hand-side of Eq.~\eqref{ddeq}, which must have rank one when the relative velocity of the DM particles is small (more concretely, when the DM relative kinetic energy is smaller than the mass splitting among the particles in the multiplet). Finally, we calculate the corresponding eigenvector to determine the enhancement factors (up to a global phase). Notice that this method cannot be  used  in the Early Universe, since pairs of charged states can exist at infinity at finite temperature, thus yielding a matrix $d^\dagger d$ with rank higher than one. 
%%%%%%%%%%%%%%%%%%%%%%%%%%%%%%%%%%%%%%%%%%%%%%%%%%%%%%
%%%%%%%%%%%%%%%
%%%%%%%%%%%%%%%%%%%%%%%%%%%%%%%%%%%%%%%%%%%%%%%%%%%%%
%%%%%%%%%%%%%%%%%%%%%%%%%%%%%%%%%%%%%%%%%%%%%%%%%%%%%
\section*{Note Added}

During the preparation of this work, we became aware of another group investigating gamma-ray signals of Minimal Dark Matter scenarios \cite{Cirelli:2015bda}, and more recently, about the analysis carried out in Ref.~\cite{Aoki:2015nza}, which also addresses the indirect detection of heavy dark matter multiplets.

%%%%%%%%%%%%%%%%%%
\vspace{0.5cm}
\section*{Acknowledgements}
We would like to thank Marco Cirelli, Thomas Hambye, Paolo Panci, Filippo Sala and  Marco Taoso for communications and Michael Gustafsson for useful discussions. This research was partially supported by the DFG cluster of excellence ``Origin and Structure of the Universe'', by the Graduiertenkolleg ``Particle Physics at the Energy Frontier of New Phenomena'' and by the TUM Graduate School. We also acknowledge support from the Belgian Federal Science Policy through the Interuniversity Attraction Pole P7/37 “Fundamental Interactions”.

%%%%%%%%%%%%%%%%%%%%%%%%%%%%%%%%%%%%%%%%%%%%%%%%%%%%%%
\bibliographystyle{apsrev.bst}
\bibliography{text}

\end{document}